\RequirePackage{fix-cm}
\documentclass[smallextended,natbib]{svjour3}

\smartqed
\usepackage{comment}
\usepackage[T1]{fontenc}
\usepackage[utf8]{inputenc}
\usepackage{lmodern}
\usepackage{amsmath,amssymb}
\usepackage{booktabs}
\usepackage{tabularx}
\usepackage{graphicx}
\usepackage{tikz}
\usepackage{pgfplots}
\usepgfplotslibrary{groupplots}
\usetikzlibrary{arrows.meta,calc,fit,patterns,positioning}
\pgfplotsset{compat=1.18}
\usepackage{xcolor}
\usepackage{url}
\usepackage[hidelinks]{hyperref}
\usepackage{microtype}
\usepackage{enumitem}

\newcolumntype{Y}{>{\raggedright\arraybackslash}X}
\newcommand{\tool}[1]{\texttt{#1}}

\journalname{Empirical Software Engineering}
\usepackage{soul}
\usepackage[most]{tcolorbox}

\tcbset{
  parskip/.style={
    before={\par\pagebreak[0]\parindent=0pt},
    after={\parfillskip=0pt\par},
  },
}
\newtcolorbox{resultbox}{%
    colback=black!3,
    colframe=black!3,
    notitle,
    sharp corners,
    borderline west={2pt}{0pt}{gray!80!black},
    enhanced,
    breakable,
    boxsep=0pt,
    left=4pt,right=2pt,top=2pt,bottom=2pt,
    before upper={\normalfont\bfseries Key Finding:\normalfont\space\itshape},
}
\begin{document}

\title{Agents on the Patch Path: Mining AI-Associated Security Remediation in GitHub Security Advisories}
\titlerunning{Mining AI-Associated Security Remediation}

\author{Claudio Segal \and Paulo Segal \and Hudson Silva Borges \and
        Paulo Silveira \and Daniel S.\ Menasch\'e \and Eduardo Santana de Almeida \and Felipe de Sant'Anna Paixão \and Joanna C. S. Santos \and Anton Kocheturov \and Gaurav Srivastava \and Raphael Machado}
\authorrunning{Segal et al.}

\institute{
Claudio Segal \and Paulo Segal \and Raphael Machado \at Fluminense Federal University (UFF),
Niter\'oi, RJ, Brazil. ORCIDs: 0009-0000-4101-4191 (C.\ Segal), 0009-0000-3265-3047 (P.\ Segal), 0000-0003-3339-9735 (R.\ Machado).
\email{csegal@id.uff.br, psegal@id.uff.br, raphaelmachado@ic.uff.br}
\and
Hudson Silva Borges \at Federal University of Mato Grosso do Sul (UFMS),
Campo Grande, MS, Brazil. ORCID: 0000-0002-8325-8773
\email{hudson.borges@ufms.br}
\and
Paulo Silveira \at Rural Federal University of Pernambuco (UFRPE),
Recife, PE, Brazil. ORCID: 0000-0003-0197-8249
\email{pauloadmsn@gmail.com}
\and
Daniel S.\ Menasch\'e \at Institute of Computing, Federal University of Rio de Janeiro
(IC--UFRJ), Rio de Janeiro, RJ, Brazil. ORCID: 0000-0002-8953-4003.
\email{sadoc@ic.ufrj.br}
\and
Eduardo Santana de~Almeida \and Felipe de Sant'Anna Paix\~ao \at Institute of Computing, Federal University of Bahia (IC-UFBA), Salvador, BA, Brazil. ORCIDs: 0000-0002-9312-6715 (E.\ S.\ de Almeida), 0009-0009-7370-5478 (F.\ S.\ Paix\~ao).
\email{esa@rise.com.br, felipe.paixao@ufba.br}
\and
Joanna C.\ S.\ Santos \at University of Notre Dame, Notre Dame, IN, USA. ORCID: 0000-0001-8743-2516.
\email{joannacss@nd.edu}
\and
Anton Kocheturov \and Gaurav Srivastava \at Siemens Corporation, Princeton, NJ, USA. ORCIDs: 0000-0003-2549-9146 (A.\ Kocheturov), 0009-0005-2755-629X (G.\ Srivastava).
\email{anton.kocheturov@siemens.com, srivastava.gaurav@siemens.com}
}

\date{Received: October 15, 2026  }

\maketitle

\begin{abstract}
\textbf{Context.} Autonomous coding agents increasingly create, review, and maintain production software.  Their participation is visible in commit identities, attribution trailers, configuration directories, and repository-level instruction files, yet little is known about their involvement in real security remediation.

\textbf{Objective.} We investigate observable evidence of coding-agent participation in commits referenced by reviewed GitHub Security Advisories (GHSAs), characterize the associated vulnerabilities, and connect these observations to the disclosure and review paths studied in prior work.

\textbf{Method.} We built a local-Git mining pipeline with four high-precision rule families: known agent accounts, explicit agent attribution in commit messages, repository instruction files, and agent configuration artifacts.  We joined the resulting commit-level findings to commit URLs appearing in GHSA references.  We distinguish direct attribution from repository-level adoption evidence, deduplicate at advisory and advisory--commit levels, and analyze tool families, severity, weakness classes, and review latency measured from the source publication.  We further introduce an exact route--timing decomposition that separates differences in review latency due to disclosure-path composition from differences within each path.

\textbf{Results.} The automated linkage identified 534 advisory--finding links across 498 advisories and 190 repositories. After excluding 29 advisories affected by human-name collisions, 469 remained across 189 repositories: 448 with direct attribution and 21 supported only by repository instruction or configuration artifacts. Claude was the most frequently observed tool family, followed by Copilot and Cursor; 200 advisories were high severity and 49 critical. Of the 469, 386 (82.3\%) were linked to GitHub Repository Advisories (GRAs).

For the 468 advisories following either the GRA-linked or NVD-first route, mean review lag was 21.95 days, compared with 19.45 days for other advisories published in the same months and 35.23 days for advisories published from July 2022 through August 2025. Against the same-month group, route proportions were similar (82.5\% versus 80.1\% GRA-linked); route composition accounted for +0.03 days of the mean difference, and within-route timing for +2.47 days (an interval including zero). Against the historical group (44.3\% GRA-linked), the mean was 13.28 days lower: -7.84 days were associated with route composition and -5.44 days with within-route timing.

\textbf{Conclusions.} Coding-agent involvement is already visible across hundreds of security advisories and repositories, including advisories addressing high- and critical-severity vulnerabilities. By distinguishing direct attribution from repository-level adoption, and route composition from within-route review timing, this study provides a clearer empirical picture of how agent-assisted remediation appears in the GHSA review pipeline.

\end{abstract}

\section{Introduction}
\label{sec:introduction}

AI coding agents are moving from code completion toward repository-level work:
they implement features, repair defects, generate tests, respond to review
comments, and create pull requests.  Large-scale datasets such as AIDev~\citep{li2025aidev} show
that these agents already operate across tens of thousands of public
repositories and leave persistent traces in software-development artifacts.  Security remediation is a particularly consequential
setting for such participation.  A successful patch must not merely compile;
it must remove the vulnerability without introducing regressions, accurately
describe the affected versions, and move through disclosure channels quickly
enough to support downstream users~\citep{siddiq2026securityageaiteammates}.

GitHub Security Advisories (GHSAs)~\citep{GitHubAdvisoryDatabase} provide an unusually useful observation
point.  A reviewed advisory includes structured severity and weakness
metadata, affected-version information, timestamps, and references that often
identify remediation commits. A recent prior work~\citep{segal2026ghsa} characterized the GHSA review
pipeline and found two dominant routes: a fast path associated with GitHub
Repository Advisories (GRAs) and a slower path in which information first
appears in the National Vulnerability Database (NVD).
That work treated patches and advisories as human-produced objects. However, the rapid
adoption of coding agents creates a new question: are AI agents visible
on the path from a security fix to a reviewed advisory?

Answering this question is harder than searching for tool names.  A commit
message may explicitly credit Claude, Copilot, Cursor, Codex, or another agent;
an exact bot identity may be present in the author or committer field; or a
repository may add an \tool{AGENTS.md}, \tool{CLAUDE.md}, or tool-specific
configuration directory. However, these evidence signals do not have identical semantics.
Explicit commit attribution is evidence that an agent participated in
producing that commit.
An instruction file is evidence of repository-level readiness or adoption, but
does not prove that an agent authored the security-relevant lines.  Moreover,
Git metadata is self-declared and names can collide with human identities.

Therefore, we study \emph{candidate AI-associated remediation} rather than
claiming verified machine authorship.  Our study first mines public
repositories associated with reviewed advisories, using a hardened local-Git
adapter and  a narrow set of rules to detect coding agents.  It then links findings to
commit URLs in advisory references.  The resulting dataset enables three
complementary analyses.  First, we quantify direct attribution and weaker
 evidence of adoption of coding agents (RQ1). Next, we characterize the security issues connected
to the candidate commits (RQ2). Finally, we relate the candidates to
advisory-source and review-time behavior (RQ3).  These analyses correspond to
the following research questions.

\begin{description}[leftmargin=1.6cm,labelwidth=1.1cm,labelsep=0.2cm,align=right,style=nextline]
\item[RQ1.] To what extent do commits referenced by reviewed GHSAs carry
observable evidence of AI coding agents, which evidence classes and tool
families account for it, and how has it grown over time?
\item[RQ2.] What is the severity, exploitability, and weakness profile of the
advisories linked to candidate AI-associated remediations, and does it
concentrate on particular vulnerability classes?
\item[RQ3.] Through which disclosure routes do candidate AI-associated
remediations enter the GHSA review pipeline, and does their review latency
differ from that of other advisories, published in the same months or in an 
earlier   period, once route composition is separated from
within-route timing?
\end{description}

Our main contributions are:

\begin{itemize}[leftmargin=1cm]
\item a reproducible, local-Git method for detecting four classes of agent
evidence in repositories associated with security advisories;
\item a linked dataset of 469 valid candidate advisories after removal of 29
advisories affected by human-name collisions, including a strict separation
between 448 direct-attribution candidates and 21 artifact-only candidates;
\item an empirical characterization of tools, severity, CWE categories, GRA
usage, and review latency in this emerging population; and
\item an exact analytical decomposition, instantiated with 12{,}951
same-month and 10{,}403 historical advisories, showing that the candidates'
review lag does not differ clearly from that of same-month advisories, and that
the lower mean relative to the historical cohort is mostly route composition
and calendar change, not faster review within a route.
\end{itemize}

The remainder of this paper is organized as follows.
Section~\ref{sec:background} reviews the GHSA review pipeline, prior work on
coding agents in real repositories, and research on security patches and
disclosure latency.  Section~\ref{sec:method} describes the study population,
the local-Git mining procedure, the detection rules, the linkage of findings to
advisories, and the validation steps.  Section~\ref{sec:results} answers the
three research questions.  Section~\ref{sec:model} develops the route--timing
decomposition that separates route composition from within-route timing.
Section~\ref{sec:discussion} interprets the evidence and its implications for
maintainers and platforms, Section~\ref{sec:threats} discusses threats to
validity, Section~\ref{sec:replication} describes the replication package,
and Section~\ref{sec:conclusion} concludes the work.

\section{Background and Related Work}
\label{sec:background}

\subsection{GitHub Security Advisories}

The GitHub Advisory Database aggregates vulnerability records from
GitHub-sourced security advisories (GHSAs), the National Vulnerability
Database (NVD), and ecosystem-specific sources. Reviewed advisories are
validated and mapped to supported package ecosystems, allowing them to
support services such as Dependabot alerts
\citep{AboutGithubAdvisoryDatabase,AboutDependabotAlets}. Advisories expose
identifiers, publication and review timestamps, severity, CVSS information,
affected packages and version ranges, CWE labels, credits, and external
references. These records therefore provide a useful basis for studying both
vulnerability characteristics and the processes through which vulnerabilities
are disclosed and reviewed.

Prior work has examined security advisories from several complementary
perspectives. Studies of vulnerability reporters and disclosure practices
investigate the roles of contributors, maintainers, and reporting platforms
in vulnerability management
\citep{alexopoulos2021vulnerability,ayala2025mixed}. Other work has examined
how maintainers review and resolve vulnerability reports and how disclosure
practices are reflected in advisory metadata
\citep{ayala2025deep,ayala2024poster,ayala2025investigating}.
Research has also investigated mechanisms for reporting security
vulnerabilities through GitHub repositories
\citep{kancharoendee2025categorizing}. Together, these studies establish the
importance of understanding not only vulnerabilities themselves but also
the people, platforms, and workflows involved in their disclosure.

Segal et al. analyzed 288,604 advisories spanning 2019--2025, including
23,563 reviewed advisories, and characterized the GHSA review pipeline
\citep{segal2026ghsa}. They found that security advisories (GHSAs) sourced from GRAs (GitHub Repository Advisories) reached review more quickly than those sourced from the NVD. Following the June
2022 NVD backfill\footnote{The 2022 NVD backfill refers to a peak surge of vulnerability metadata updates on GHSA database that year. 
Prior to december 2021, NVD entries were manually imported into the GHSA database before transitioning to an automated ingestion pipeline~\citep{GitHubBlogNVDBackfill}. Thus, a comprehensive backfill was executed in June 2022 to retroactively incorporate all historical NVD advisories into the repository. }, publication-to-review delays became short for both
source groups, whereas the difference in patch-to-review latency remained
substantial. Their model characterized these patterns through two paths: a
direct path for GRA-first advisories and an NVD-first path involving an
additional stochastic delay. 
While this prior work also examined GHSAs, it
did not mine repository histories for coding-agent evidence, link such evidence
to advisory commits, characterize AI-associated remediation, or provide the
route--timing decomposition introduced in our work.
Thus, our study complements the pipeline-level analysis performed by Segal et al. \citep{segal2026ghsa} by changing the
unit of interest. Rather than focusing exclusively on advisory sources and
review latency, we begin with commits referenced by security advisories and
investigate whether repository history contains explicit evidence of coding
agent involvement. This perspective connects the disclosure process to the
development activity underlying the security fixes.

\subsection{AI Coding Agents and Software Security}

Recent coding agents can generate and modify code, interact with development
tools, and contribute to repository workflows. Their growing adoption has
motivated empirical studies of both the security properties of generated
code and the characteristics of agent contributions in real-world projects.

One line of research investigates the security of AI-generated code.
Pearce et al. examined the security of code contributions generated by GitHub
Copilot \citep{pearce2025asleep}, while Fu et al. conducted an empirical study
of security weaknesses in Copilot-generated code in GitHub projects
\citep{fu2025securityweaknessescopilotgeneratedcode}. These studies motivate
continued scrutiny of AI-generated software, particularly when generated
code is incorporated into production repositories. However, the presence of
potential security weaknesses in generated code is distinct from evidence
that a coding agent participated in a particular vulnerability fix. Establishing
that connection requires examining development artifacts and their provenance.

A complementary line of research studies coding agents as participants in
software development. Li et al. introduced the AIDev dataset to characterize
agentic coding activity in GitHub pull requests, including contributions from
tools such as Copilot, Devin, Claude Code, Cursor, and Codex
\citep{li2025aidev}. More recently, Siddiq et al. investigated the security
of agentic pull requests on GitHub, bringing the security implications of
agent participation in repository workflows into focus
\citep{siddiq2026securityageaiteammates}. This body of work moves beyond
synthetic coding benchmarks to examine agent activity in real development
settings.

Closer to the vulnerability-disclosure setting, Rooijendijk et al.\ compared
how humans, bots, and autonomous agents mention vulnerability identifiers
(CVE, CWE, GHSA, and others) in GitHub pull requests, using AIDev and an
augmented set of pull requests from the same repositories
\citep{rooijendijk2026said}.  They found that bots account for most
mentions, mainly in dependency-update and audit descriptions, whereas human
and agent mentions are rarer but appear in more places, such as commit
messages and titles, where they support fixes and discussion. Their unit of
analysis is the identifier mention in pull-request text.

Our approach is complementary to these studies in both its sampling strategy
and its unit of analysis. Rather than beginning with a catalog of agent
pull requests and subsequently examining their properties, we select GHSAs  and the commits they reference. We then investigate whether the
associated repository history contains explicit agent evidence. Depending on
the available artifacts, this can include direct commit attribution,
attribution retained in merged or squashed commits, Copilot Autofix markers,
and repository-level instruction or configuration files.

These forms of evidence are not equivalent. Direct attribution can provide
evidence of an agent's participation in a particular change, whereas
repository-level configuration artifacts establish a weaker connection to
the repository's development environment. Accordingly, identifying an
agent-associated repository is not sufficient to conclude that an agent
produced a particular security patch. By distinguishing evidence types, our
study seeks to characterize the observable relationship between coding
agents and security advisories without treating agent adoption as proof of
agent authorship.

\subsection{Security Patches and Vulnerability Disclosure Latency}

Security patches differ from ordinary software changes in their complexity,
review requirements, and coordination with releases
\citep{li2017large}. Research on vulnerability management has examined
vulnerability disclosure, patch availability, dependency exposure, and
downstream adoption. For example, Arora et al. investigated software vendors'
patch-release behavior in relation to vulnerability disclosure
\citep{arora2010empirical}; Ponta et al. studied the detection, assessment,
and mitigation of vulnerabilities in open-source dependencies
\citep{ponta2020detection}; and Decan et al. analyzed the impact of
vulnerabilities in the npm dependency network
\citep{decan2018impact}. Related work has examined the security risks
associated with dependency ecosystems
\citep{zimmermann2019small}.

The timing of security patches and disclosures is crucial because
publicly available patches may reveal vulnerability details before all
affected users have become aware of, or adopted, a fix
\citep{wang2017characterizing}. Research on vulnerability disclosure
management has also examined the factors influencing disclosure and
remediation in open-source projects
\citep{liu2025empirical}. These studies motivate treating the publication,
patching, and review of an advisory as distinct events rather than as a
single measure of vulnerability-handling speed.

Automation can influence several stages of this process. Research on
automated dependency updates has investigated how tools such as Dependabot
affect developer workflows and dependency-update behavior
\citep{mirhosseini2017can,he2023automating}. More broadly, studies of
GitHub Actions have examined how developers use automation to coordinate
repository workflows \citep{kinsman2021software}. Prior work on vulnerability
databases and security-event data provides additional context for the
consistency and timeliness of vulnerability information across sources
\citep{bozorgi2010beyond,allodi2017security}. These studies are relevant to
the present work because the path through which an advisory enters the GHSA
database can affect when it becomes visible and when it is reviewed.

Coding agents introduce additional possible mechanisms. Agent participation
could help developers produce patches, tests, or disclosure materials more
quickly. Conversely, maintainers may need additional review or revision
cycles for agent-associated security changes. Existing work on generated-code
security and agentic pull requests motivates both possibilities, but does not
by itself establish how agent involvement relates to the review latency of
security advisories.

This study therefore distinguishes disclosure route from review timing and
examines agent evidence in the commits referenced by advisories. Its focus is
not to assume that agent involvement accelerates or delays security
remediation, but to investigate whether explicit evidence of such involvement
is associated with observable patterns in the advisory review process.

\section{Study Methodology}
\label{sec:method}

\begin{figure}[!h]
    \centering
    \includegraphics[width=\linewidth]{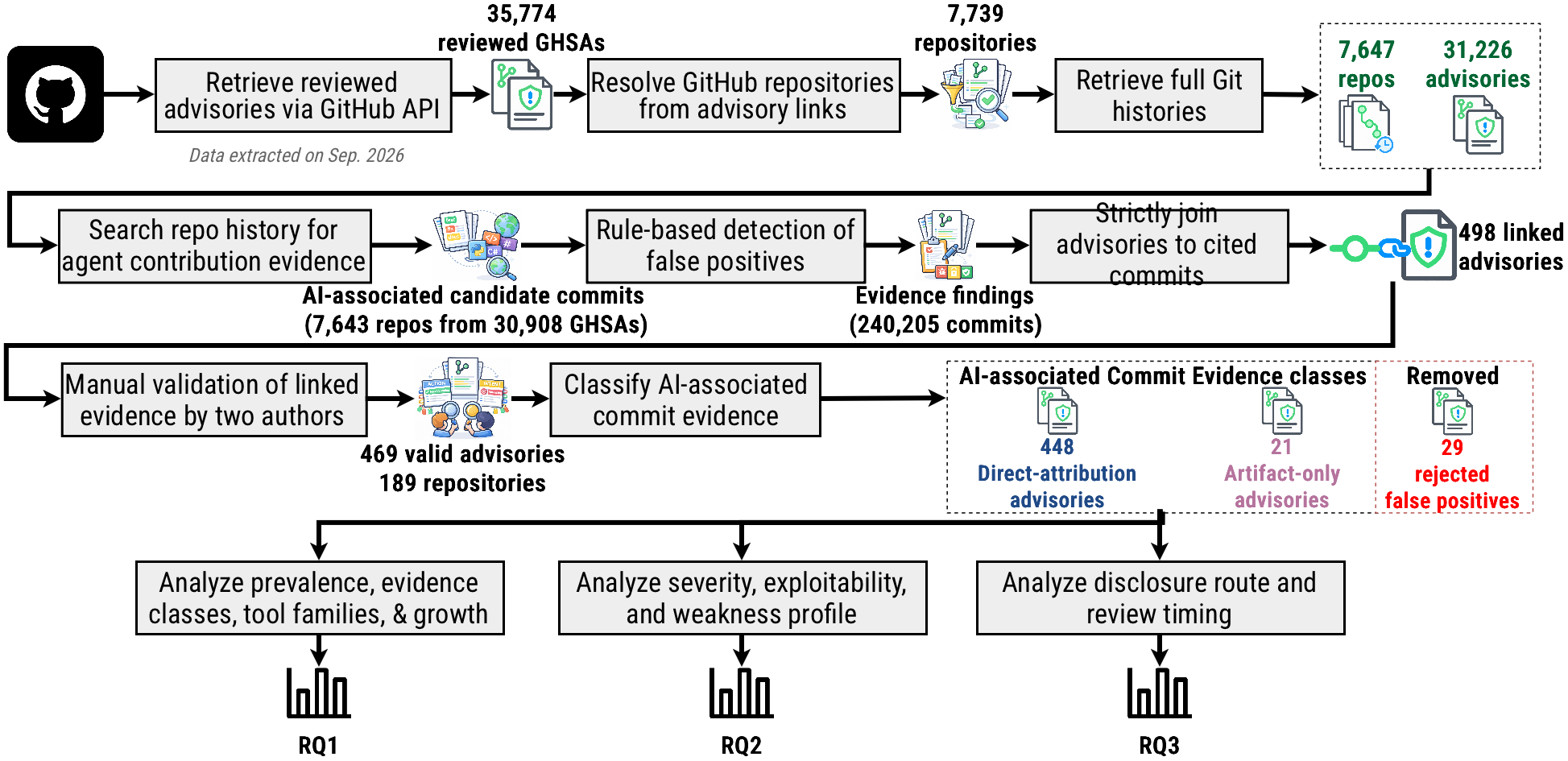}
    \caption{Overview of our study's methodology.}
    \label{fig:overview}
\end{figure}

Figure~\ref{fig:overview} shows an overview of our methodology, which is elaborated in details in the next subsections.

\subsection{Retrieving Reviewed Advisories}

We collected every \textit{reviewed advisory} in the GitHub Advisory Database through
GitHub's public API on September 23, 2026 obtaining a total of 35,774 advisories.  We restrict the study to reviewed
advisories because only these have been curated by GitHub: they are mapped to
package ecosystems and affected versions, drive Dependabot alerts, and carry
the review timestamp that RQ3 requires. For each advisory, we stored the complete record returned by the API. Our analyses use the source code location and references for repository linkage (Section~\ref{sub:linking-advisories}); the severity, CVSS, EPSS, and CWE fields for the security profile (RQ2); and the publication, GitHub review, and NVD publication timestamps, together with the repository-advisory URL that identifies the disclosure route, for the timing analysis (RQ3).

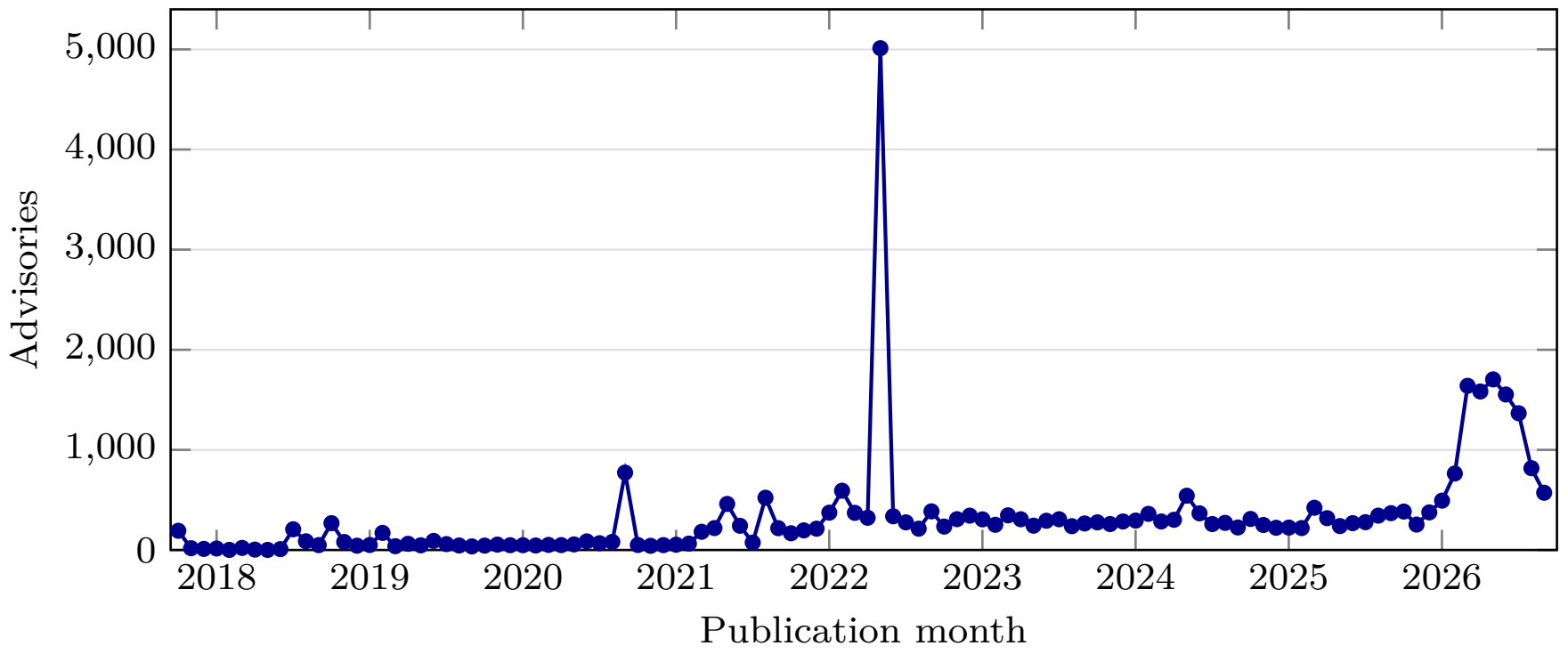
\begin{figure}[!ht]
\centering
\begin{tikzpicture}
\begin{axis}[
  width=\textwidth,
  height=5.6cm,
  xmin=2017.70,
  xmax=2026.75,
  ymin=0,
  ymax=5400,
  xlabel={Publication month},
  ylabel={Advisories},
  xtick={2018,2019,...,2026},
  x tick label style={/pgf/number format/.cd,1000 sep={},precision=0,fixed},
  ytick={0,1000,...,5000},
  y tick label style={/pgf/number format/1000 sep={\text{,}}},
  scaled y ticks=false,
  ymajorgrids,
  grid style={gray!25},
  axis line style={line width=0.5pt},
  tick style={line width=0.5pt},
  tick label style={font=\small},
  label style={font=\small},
]
\addplot[blue!55!black,line width=0.8pt,mark=*,mark size=1.3pt]
coordinates {
(2017.7500,192) (2017.8333,18) (2017.9167,11) (2018.0000,15) (2018.0833,1) (2018.1667,21)
(2018.2500,4) (2018.3333,1) (2018.4167,9) (2018.5000,207) (2018.5833,87) (2018.6667,49)
(2018.7500,268) (2018.8333,80) (2018.9167,43) (2019.0000,50) (2019.0833,171) (2019.1667,38)
(2019.2500,62) (2019.3333,47) (2019.4167,91) (2019.5000,59) (2019.5833,45) (2019.6667,36)
(2019.7500,44) (2019.8333,54) (2019.9167,48) (2020.0000,48) (2020.0833,45) (2020.1667,51)
(2020.2500,49) (2020.3333,56) (2020.4167,86) (2020.5000,68) (2020.5833,81) (2020.6667,773)
(2020.7500,50) (2020.8333,42) (2020.9167,49) (2021.0000,53) (2021.0833,64) (2021.1667,182)
(2021.2500,219) (2021.3333,460) (2021.4167,242) (2021.5000,74) (2021.5833,522) (2021.6667,218)
(2021.7500,167) (2021.8333,196) (2021.9167,213) (2022.0000,373) (2022.0833,592) (2022.1667,371)
(2022.2500,322) (2022.3333,5012) (2022.4167,337) (2022.5000,276) (2022.5833,213) (2022.6667,386)
(2022.7500,234) (2022.8333,308) (2022.9167,343) (2023.0000,306) (2023.0833,252) (2023.1667,346)
(2023.2500,307) (2023.3333,243) (2023.4167,292) (2023.5000,307) (2023.5833,238) (2023.6667,265)
(2023.7500,276) (2023.8333,259) (2023.9167,285) (2024.0000,294) (2024.0833,361) (2024.1667,285)
(2024.2500,302) (2024.3333,542) (2024.4167,366) (2024.5000,259) (2024.5833,271) (2024.6667,225)
(2024.7500,310) (2024.8333,250) (2024.9167,222) (2025.0000,224) (2025.0833,219) (2025.1667,422)
(2025.2500,318) (2025.3333,239) (2025.4167,268) (2025.5000,278) (2025.5833,343) (2025.6667,368)
(2025.7500,384) (2025.8333,254) (2025.9167,376) (2026.0000,493) (2026.0833,764) (2026.1667,1640)
(2026.2500,1583) (2026.3333,1704) (2026.4167,1553) (2026.5000,1366) (2026.5833,817) (2026.6667,572)
};
\end{axis}
\end{tikzpicture}
\caption{Reviewed advisories published per month in the study population
($n=35{,}774$). September~2026 covers only the days before the collection
date. The spike around June 2022 corresponds to the NVD backfill.}
\label{fig:population-monthly}
\end{figure}

Figure~\ref{fig:population-monthly} shows the monthly publication volume of
the reviewed advisories, which span October~2017 to September~2026 (the last
month is partial).  The isolated spike in 2022 reflects GitHub’s automated backfill of historical NVD advisories into GHSA, as
explained in Section~\ref{sec:background}. The volume of advisories also increases sharply in 2026:
the 2026 portion of the snapshot contains 10,492 advisories, compared with
3,376--3,693 per year in 2023--2025.  The growth is broad rather than driven by
a single project; \tool{openclaw/openclaw}, the largest contributor, accounts
for 629 of these advisories.  Since most candidate advisories were also
published in 2026 (Section~\ref{sec:results}), calendar time is a potential
confounder in any comparison with earlier advisories
(Section~\ref{sec:threats}).

\subsection{Linking Advisories to GitHub Repositories}
\label{sub:linking-advisories}
We then identified the repositories associated with each advisory,
verified their availability on GitHub, and collected their metadata.
Specifically, repositories were identified primarily from the advisory's \textit{source code
location} field.  When this field was absent or did not resolve to a GitHub
repository, we took the owner and repository from GitHub commit URLs in the
advisory's external references.

After analyzing the advisories' metadata, we were able to locate the corresponding GitHub repository for  31,343 out of 35,774 advisories, resulting in a total of 7,739 distinct repositories; the remaining 4,431 GHSAs have neither a valid GitHub source code location nor a commit reference to a single GitHub repository.  
It is important to note that, at repository metadata collection time, 92 of these 7,739 repositories were no longer publicly available. 
Excluding the 117 advisories associated with them leaves our dataset  with a total of 31,226 advisories linked to the remaining 7,647 repositories, which belong to 5,019 distinct owners (3,125
organizations and 1,894 individual users).

\begin{figure}[!ht]
\centering
\begin{tikzpicture}
\begin{axis}[
  width=0.96\textwidth,
  height=6.6cm,
  ybar,
  bar width=13pt,
  ymin=0,
  ymax=5200,
  xlabel={Advisories per repository},
  ylabel={Repositories},
  symbolic x coords={1,2,3,4,5,6--10,11--20,21--50,51--100,101+},
  xtick=data,
  ymajorgrids,
  grid style={gray!25},
  axis line style={line width=0.5pt},
  tick style={line width=0.5pt},
  tick label style={font=\small},
  x tick label style={font=\footnotesize},
  label style={font=\small},
  scaled y ticks=false,
  y tick label style={/pgf/number format/1000 sep={{{,}}}},
  nodes near coords,
  every node near coord/.append style={font=\footnotesize,text=black,/pgf/number format/1000 sep={{{,}}}},
  enlarge x limits=0.06,
]
\addplot+[draw=black!80,fill=blue!55!black,postaction={pattern=north east lines},line width=0.6pt]
coordinates {(1,4557) (2,1170) (3,511) (4,325) (5,183) (6--10,417) (11--20,238) (21--50,138) (51--100,38) (101+,27)};
\end{axis}
\end{tikzpicture}
\caption{Distribution of reviewed advisories per GitHub repository, for the
repositories identified through a valid source code location. Advisories
without a valid GitHub location are excluded}
\label{fig:population-per-repo}
\end{figure}
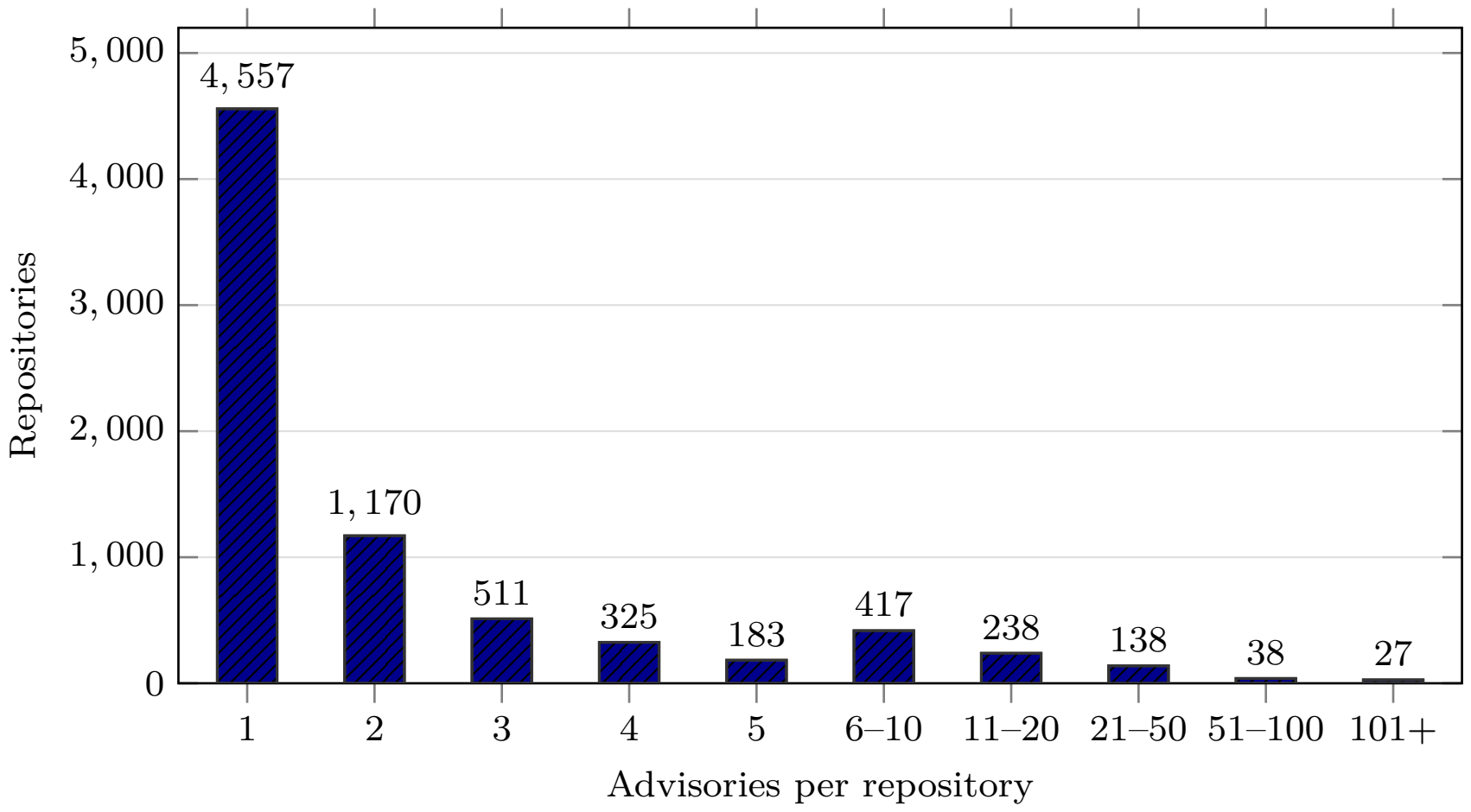

Figure~\ref{fig:population-per-repo} shows how advisories are distributed
across the repositories identified through a valid source code location.
The distribution is strongly right-skewed: most repositories have a single
advisory, whereas a small tail of projects, led by \tool{openclaw/openclaw}
(629 advisories) and \tool{tensorflow/tensorflow} (421), contributes many.
Because of this concentration, our analyses deduplicate at the advisory level
and use repository-cluster resampling when estimating uncertainty
(Section~\ref{sec:decomposition}).

\subsection{Mining repository histories for agent evidence}
\label{sec:evidence-search}

\begingroup\sloppy
After obtaining advisory metadata and finding their corresponding GitHub repo, in this stage, we mine each of these repositories histories.  Our goal in this step is to searches each
repository's version history for traces of AI coding-agent use, and
Section~\ref{sec:linking} later links the resulting findings to the commits
that advisories cite.  We were able to fully mine 7,643 out of the 7,647 available repositories. The remaining four repositories \footnote{The repositories we were unable to mine were: \path{liferay/liferay-portal},
\path{hashicorp/terraform-provider-aws}, \path{home-assistant/core}, and
\path{NousResearch/hermes-agent}.} presented problems during collection, and
we could not obtain their data.  Our results therefore do not cover these
repositories or the 318 advisories (0.9\% of the population) associated with
them, 297 of which belong to \path{liferay/liferay-portal}.

For each of the 7,643 repositories associated with 30,908 advisories, we performed the following:\par\endgroup

\begin{enumerate}[leftmargin=*]
\item \textbf{Acquisition:}  We cloned each repository with its full commit
history and pinned the head of its default branch.  Thus, all subsequent search queries
operate on the history reachable from that commit.  

\item \textbf{Broad candidate search:} In this step, 
we performed a  keyword-based search over the commit's metadata in order to identify \textit{candidate  commits} that might carry agent
evidence. Specifically, each repository was queried with a filtered \textit{git log} command over the pinned history, using fixed lists of terms and paths (Appendix~\ref{app:catalog}). The search keywords were matched as literal, case-insensitive substrings, without regular expressions or learned classifiers. The search query was deliberately designed to favor recall and covered four types of commit metadata:
  \begin{enumerate}[label=(\alph*)]
  \item names and e-mail addresses in the author and committer fields;
  \item authorship and co-authorship references in commit messages, including
  \tool{Co-authored-by} trailers;
  \item additions or renames of agent instruction files at the repository root
  (e.g., \tool{AGENTS.md}, \tool{CLAUDE.md}); and
  \item additions or modifications of agent configuration files (e.g., files
  under \tool{.claude}, \tool{.cursor}, or \tool{.codex}).
  \end{enumerate}
    \item \textbf{Deduplication}: In this last step, we  remove duplicated commits that were matched independently across the four
searches.
\end{enumerate}
 It is important to note that since the keyword search was broad, and favored recall over precision, in the next step (Section~\ref{sec:rules}) we  employ deterministic rule-based detection rules to discard false positives from this list of \textit{candidate commits}. This resulted in a total of 240,205 \textit{distinct commits}.

\subsection{Rule-based Detection of Commits that are False Positives}
\label{sec:rules}
\begingroup\sloppy

The candidate commits previously collected may include false positives for many reasons. For example, 
Git metadata is not cryptographically authenticated: a contributor can write an
arbitrary co-author trailer, and a tool-like name can belong to a human.  Thus, we
 treat every signal as a \emph{declaration} and use \textit{``direct
attribution''} rather than \textit{``verified authorship''}. 
Consequently, a
candidate commit is only retained in our study if it is confirmed by at least one of the four deterministic, precision-oriented rules
shown in Table~\ref{tab:rules}.  Two rules read the commit message and identity
fields; the other two read the repository paths that the commit changes.  The
complete catalog of recognized names, handles, and paths is given in
Appendix~\ref{app:catalog}.

\begin{table*}[ht]
\caption{Active detection rules and the claims they can support.}
\label{tab:rules}
\centering
\footnotesize
\begin{tabularx}{\textwidth}{@{}p{2.1cm}p{2.2cm}Y Y@{}}
\toprule
\textbf{Rule} & \textbf{Category} & \textbf{Evidence boundary} & \textbf{Interpretation in this study} \\
\midrule
Message signal & Direct attribution & Explicit attribution in the complete
commit message: a well-formed co-author trailer naming a recognized agent,
an explicit generation phrase, or a distinctive Copilot workflow phrase &
Declared agent participation in that commit \\
Known bot identity & Direct attribution & Author or committer name exactly
equal to an allowlisted AI-bot handle; co-author trailers excluded &
Declared agent participation in that commit \\
Instruction file & Adoption artifact & Addition or rename of a recognized,
case-sensitive agent-instruction file at the repository root, such as
\tool{AGENTS.md} or \tool{CLAUDE.md} & The commit introduced agent-facing
instructions; not commit authorship \\
Agent configuration & Adoption artifact & Addition or rename of a recognized
root configuration file, or addition of a file below a recognized top-level
directory such as \tool{.claude}, \tool{.cursor}, or \tool{.codex} &
The commit configured the repository for an agent tool; not tool execution or
commit authorship \\
\bottomrule
\end{tabularx}
\end{table*}

\paragraph{Commit-message attribution.}
The message rule accepts three kinds of signal.  The first is a well-formed,
line-anchored \tool{Co-authored-by} trailer whose name is an allowlisted bot
handle or contains a recognized tool name as a whole token, or whose address is
one of the canonical Copilot, Anthropic, or OpenAI addresses.  The second is an
explicit generation phrase such as ``generated by GitHub Copilot'' or
``AI-generated.''  The third is a distinctive workflow form such as a
standalone ``Apply Copilot suggestion'' line or ``Copilot Autofix.''

\paragraph{Known bot identity.}
The identity rule requires the author or committer name to equal,
case-insensitively, an allowlisted agent account such as
\tool{copilot-swe-agent[bot]} or \tool{devin-ai-integration[bot]}.  It does not
verify the account against GitHub, and it ignores co-author trailers: a
human-authored commit that lists an AI co-author is handled by the message rule
rather than classified as bot-authored.

\paragraph{Adoption artifacts.}
The two path rules detect the introduction of agent-facing files: an
instruction file at the repository root, such as \tool{AGENTS.md} or
\tool{CLAUDE.md}, or project-level configuration, either a recognized root file
(e.g., \tool{.cursorrules}) or any file below a recognized top-level directory
(e.g., \tool{.claude} or \tool{.cursor}).  Additions and renames into these
paths count, names are matched exactly and case-sensitively, and file contents
are not inspected.  A path finding therefore marks the commit that
\emph{introduced} an artifact, not the repository's state at later commits.
This distinction matters when findings are linked to advisories
(Section~\ref{sec:linking}).

\paragraph{Visibility of attribution.}
Attribution is observable only if it survives the project's merge workflow.
When a pull request is squashed or rebased, the commit that reaches the default
branch carries only the metadata and message that the merge produces;
depending on how that message is composed, co-author trailers from the original
commits may be preserved or lost.  The rules see only the commits that reach
the pinned history, so the corpus reflects attribution that survives these
workflows.

\endgroup

At the end of this step, we retained a total of 240,205 commits.

\subsection{Linking evidence findings to security advisories}
\label{sec:linking}

The previous steps produced evidence findings at the commit level,
independently of any advisory.  This linking step ties those findings to changes
directly related to resolving a vulnerability.  We therefore linked evidence
and advisories through the commits that the advisories themselves cite, rather
than through repository membership alone: a repository that uses an agent
somewhere in its history is not, by that fact, evidence that an agent took part
in a security fix.

Advisory references are the natural anchor for this association: when a fix is
public, a reviewed GHSA commonly lists the URL of the remediation commit.  For
each advisory, we extracted the repository and commit identifier from every
GitHub reference whose path contains a \tool{/commit/} segment, and we linked
the advisory to every finding recorded for that repository and commit.  Two
constraints keep the join strict.  Only full 40- or 64-character commit
identifiers are used, because abbreviated identifiers can be ambiguous.  The
repository must match as well as the identifier, because forks share commit
objects with their upstream, so an identifier-only join could attach a finding
from one repository to an advisory about another.  Advisories without commit
references, or whose referenced commits have no findings, are outside the
candidate set.

The result is a many-to-many relation:  an advisory can cite several commits
(for example, a fix and its backports), a commit can be cited by several
advisories, and one commit can carry several findings.  We therefore report
counts at each level of aggregation---linked rows, advisory--commit pairs,
distinct commits, and distinct advisories and repositories
(Table~\ref{tab:corpus})---and advisory-level summaries use one record per GHSA
identifier.

An advisory is classified as \emph{direct attribution} if at least one of its
linked findings comes from the message or identity rule, and as
\emph{artifact only} if all of its linked findings come from the path rules.
The second class is narrower than its name suggests.  Because a path finding
marks the commit that introduced an artifact, an advisory is artifact-only when
the cited commit itself added an instruction or configuration file; a fix made
in a repository that already contained \tool{AGENTS.md} produces no artifact
link.  Artifact-only advisories therefore capture agent configuration arriving
in the same change as the remediation, not repository-level adoption in
general.
Tool-family counts are multi-label, because the findings linked to one advisory
can credit more than one agent.

\subsection{Validation of trace links between commits and advisories}
\label{sec:validation}

The advisories linked to candidate commits were manually validated by two authors
of this paper, who checked whether the evidence matched in each linked commit
supports the signal assigned by the detection rule. 
After validation, 469 out of the 498 linked advisories  were accepted.
All 29 rejections were human-name collisions, i.e.,  the linked commits carried
co-author trailers naming one of two human contributors whose first name is
Devin, which the message rule had matched as the Devin agent.  One collision
affected a single advisory; the other, a contributor to a single repository, affected
32 advisories.  In 28 of them the trailer was the only evidence, and we removed
those advisories before classification (Section~\ref{sec:results}).  In the
remaining four, the same commits also carried an explicit ``AI-assisted''
statement; we disregarded the trailer and retained the advisories on the basis
of that statement alone.  The case shows that even distinctive tool names can
collide with personal names, and that a single prolific contributor can
multiply one collision across many advisories, so tool name  matches in
co-author trailers require validation rather than being accepted as
attribution by construction.
\subsection{Study Dataset Overview}

Table~\ref{tab:corpus} presents the linkage counts in  our corpus used in our study. The collected advisories cite 25{,}956
GitHub commit URLs, of which 25{,}601 carry a parseable repository and full
commit identifier.  The strict repository-and-identifier join yields 534
advisory--finding links.  A further 11 identifier matches were rejected
because the cited repository differed from the mined one, mostly renamed or
transferred repositories and fixes cited from personal forks.  The links form 525 advisory--commit pairs, because
one commit can carry more than one finding, and refer to 440 distinct commits
and 498 advisories across 190 repositories.  Validation removed 29 advisories
whose only evidence was a co-author trailer naming a human contributor called
Devin (Section~\ref{sec:validation}), leaving 469 valid candidate advisories
across 189 repositories.

\begin{table}[t]
\caption{Candidate-corpus construction}
\label{tab:corpus}
\centering
\begin{tabular}{lr}
\toprule
Measure & Count \\
\midrule
Advisory references to GitHub commits & 25{,}956 \\
\quad with parseable repository and full identifier & 25{,}601 \\
Advisory--finding links (repository and identifier match) & 534 \\
Distinct advisory--commit pairs & 525 \\
Distinct linked commits & 440 \\
Valid candidate advisories & 469 \\
\quad Direct attribution & 448 \\
\quad Artifact only & 21 \\
Distinct repositories & 189 \\
\bottomrule
\end{tabular}
\end{table}
\section{Empirical Results}
\label{sec:results}

\subsection{RQ1: Prevalence, evidence classes, tool families, and growth}

We found that, of the 469 valid candidate advisories, 448 (95.5\%) of them (spanning 183 repositories) contain at
least one direct-attribution signal, while 21 (4.5\%), from eight
repositories, are supported only by an instruction or configuration artifact.
At the advisory--commit level, the corresponding counts are 473
direct-attribution pairs and 21 artifact-only pairs.  Direct attribution comes
almost entirely from commit messages: the message rule supports 447 advisories
and the bot-identity rule four.  This asymmetry matters:
the principal empirical result concerns direct public attribution, whereas the
artifact-only subset is evidence that the repository had begun configuring an
agent-oriented workflow.

Tool-family detection is multi-label and is computed over direct-attribution
evidence.  Claude appears in 306 of the 448 direct-attribution advisories
(68.3\%), Copilot in 57, Cursor in 42, Codex in 12, Gemini in four, Devin in
three, CodeRabbit in two, and Aider and Sourcery in one each.  A further 31
advisories carry only a generic phrase such as ``AI-assisted'' that names no
tool.  These counts should not be interpreted as market share.  Tools differ in
how consistently they emit trailers, and the rule catalog intentionally favors
signals with distinctive names and canonical identities.

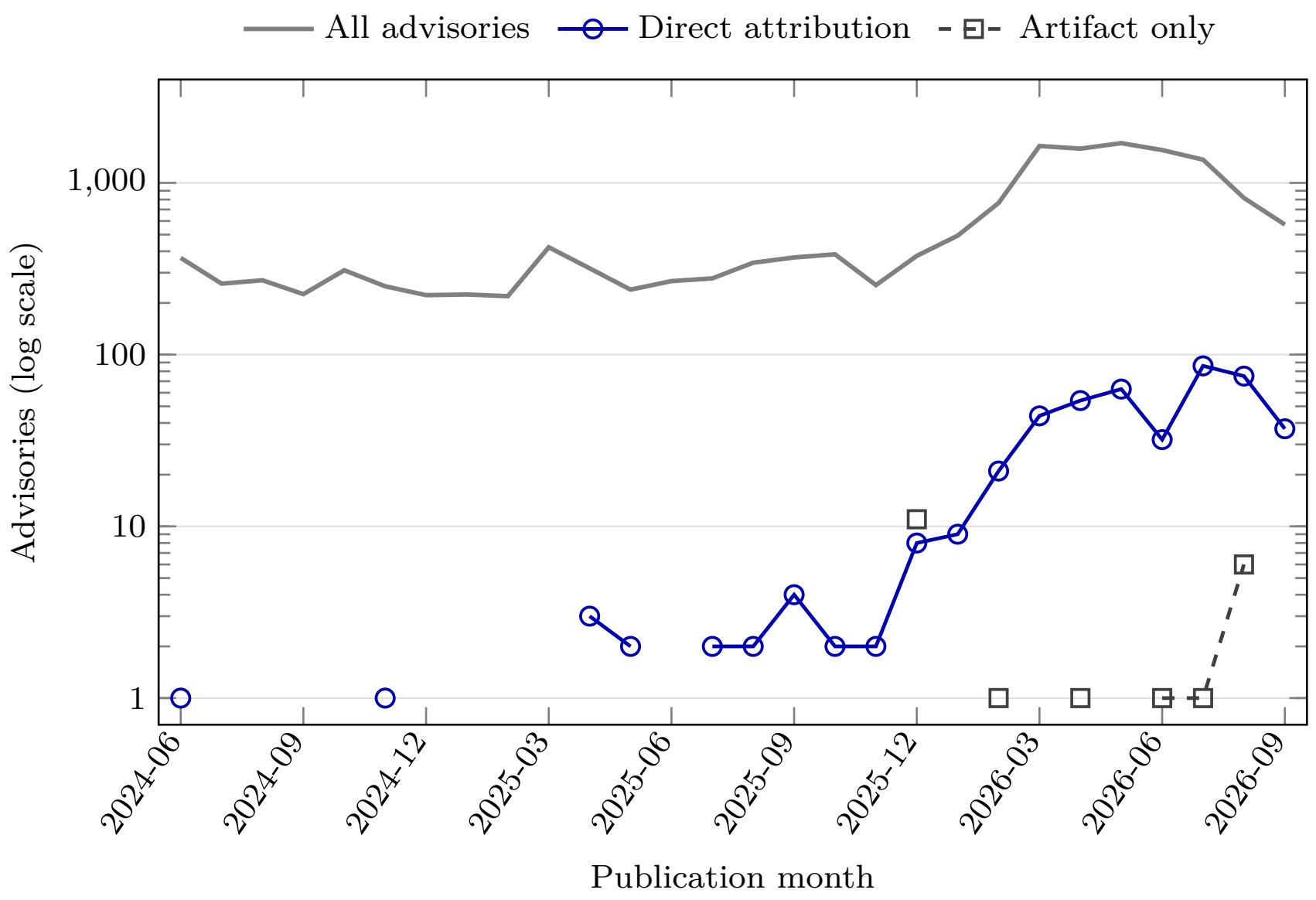
\begin{figure}[t]
\centering
\begin{tikzpicture}
\begin{axis}[
  width=0.96\textwidth,
  height=7.1cm,
  xlabel={Publication month},
  ylabel={Advisories (log scale)},
  ymode=log,
  ymin=0.7, ymax=4000,
  ytick={1,10,100,1000},
  yticklabels={1,10,100,1{,}000},
  ymajorgrids,
  grid style={gray!25},
  axis line style={line width=0.5pt},
  tick style={line width=0.5pt},
  symbolic x coords={2024-06,2024-07,2024-08,2024-09,2024-10,2024-11,2024-12,2025-01,2025-02,2025-03,2025-04,2025-05,2025-06,2025-07,2025-08,2025-09,2025-10,2025-11,2025-12,2026-01,2026-02,2026-03,2026-04,2026-05,2026-06,2026-07,2026-08,2026-09},
  xtick={2024-06,2024-09,2024-12,2025-03,2025-06,2025-09,2025-12,2026-03,2026-06,2026-09},
  enlarge x limits=0.02,
  unbounded coords=jump,
  x tick label style={rotate=55,anchor=east,font=\small},
  y tick label style={font=\small},
  label style={font=\small},
  legend style={at={(0.5,1.03)},anchor=south,draw=none,fill=none,font=\small,/tikz/every even column/.append style={column sep=0.5em}},
  legend columns=3,
]
\addplot+[black!50,solid,line width=1.1pt,mark=none]
coordinates {(2024-06,366) (2024-07,259) (2024-08,271) (2024-09,225) (2024-10,310) (2024-11,250) (2024-12,222) (2025-01,224) (2025-02,219) (2025-03,422) (2025-04,318) (2025-05,239) (2025-06,268) (2025-07,278) (2025-08,343) (2025-09,368) (2025-10,384) (2025-11,254) (2025-12,376) (2026-01,493) (2026-02,764) (2026-03,1640) (2026-04,1583) (2026-05,1704) (2026-06,1553) (2026-07,1366) (2026-08,817) (2026-09,572)};
\addlegendentry{All advisories}
\addplot+[blue!70!black,solid,line width=0.9pt,mark=o,mark size=2.2pt,mark options={fill=white,line width=0.7pt}]
coordinates {(2024-06,1) (2024-07,nan) (2024-08,nan) (2024-09,nan) (2024-10,nan) (2024-11,1) (2024-12,nan) (2025-01,nan) (2025-02,nan) (2025-03,nan) (2025-04,3) (2025-05,2) (2025-06,nan) (2025-07,2) (2025-08,2) (2025-09,4) (2025-10,2) (2025-11,2) (2025-12,8) (2026-01,9) (2026-02,21) (2026-03,44) (2026-04,54) (2026-05,63) (2026-06,32) (2026-07,86) (2026-08,75) (2026-09,37)};
\addlegendentry{Direct attribution}
\addplot+[black!75,dashed,line width=0.9pt,mark=square,mark size=2.1pt,mark options={fill=white,line width=0.7pt,solid}]
coordinates {(2024-06,nan) (2024-07,nan) (2024-08,nan) (2024-09,nan) (2024-10,nan) (2024-11,nan) (2024-12,nan) (2025-01,nan) (2025-02,nan) (2025-03,nan) (2025-04,nan) (2025-05,nan) (2025-06,nan) (2025-07,nan) (2025-08,nan) (2025-09,nan) (2025-10,nan) (2025-11,nan) (2025-12,11) (2026-01,nan) (2026-02,1) (2026-03,nan) (2026-04,1) (2026-05,nan) (2026-06,1) (2026-07,1) (2026-08,6) (2026-09,nan)};
\addlegendentry{Artifact only}
\end{axis}
\end{tikzpicture}
\caption{Monthly candidate advisories by evidence class and all reviewed
advisories published in the same month, on a logarithmic scale. Direct
attribution is shown with solid circles and artifact-only evidence with dashed
squares; months without candidates in a class have no marker.
September~2026 covers only the days before the collection date}
\label{fig:monthly}
\end{figure}

Figure~\ref{fig:monthly} compares the monthly candidate counts with all
advisories published in the same months.  Of the 469 valid advisories, two were
published in 2024, 36 in 2025, and 431 in 2026 (through September).  This growth
does not merely follow the volume of advisories.  The monthly total also rose,
from at most 422 in 2025 to between 1{,}366 and 1{,}704 from March to July
2026, but the candidates' share of all advisories rose faster: from 0.05\% in
2024 to 1.0\% in 2025 and 4.1\% in 2026.  Monthly, the share stayed at or below
1.1\% until November 2025, ranged from 1.8\% to 3.7\% in the first half of
2026, and reached 6.4\% to 9.9\% from July to September 2026.  The isolated
peak of December 2025 (5.1\%) is mostly artifact-only evidence (11 of 19
candidates).  The concentration is consistent with recent agent adoption, but it also
creates a strong calendar-time confound for any comparison with historical
advisories.

\begin{resultbox}
Public histories provide direct agent-attribution evidence for 448 reviewed
security advisories across 183 repositories.  Repository artifacts identify a
further 21 advisories, but those observations must not be presented as
agent-authored security fixes.
\end{resultbox}

\subsection{RQ2: Severity, exploitability, and weakness profile}
\label{sec:rq2}

Table~\ref{tab:security} profiles the 469 valid candidates against a
month-standardized reference: the 13{,}199 other reviewed advisories published
in the 19 months that contain candidates, reweighted so that each month has the
weight it has among the candidates.  Because 92\% of the candidates were
published in 2026, this reference keeps time changes in the advisory
population from being read as a property of the candidates.

Severity closely follows the reference.  High-severity advisories are the
largest class (200, 42.6\%), followed by medium (193, 41.2\%), critical (49,
10.4\%), and low (27, 5.8\%); the reference shares are 39.4\%, 41.5\%,
10.9\%, and 8.2\%.  Nearly all candidates carry a CVSS base score (462,
98.5\%): 350 have a CVSS~v3 score, with a median of 7.4 (reference 7.1) and a
mean of 6.97, and a further 112 have only a CVSS~v4 score.  EPSS values are
available for 369 candidates; their median is 0.0035 (reference 0.0034), and
only two exceed 0.1, although the maximum is 0.82.  The same contrast between
high severity and low exploitation probability holds in the reference, so it
reflects the advisory population rather than a trait of the candidates.  The
share of GRA-linked advisories (82.3\% versus 79.6\%) is also close to the
reference.  The share reviewed within 24 hours of first disclosure, measured as
in Section~\ref{sec:rq3}, is somewhat lower for the candidates (16.5\% of 468
versus 20.9\%); Section~\ref{sec:rq3} examines disclosure routes and review
timing.

\begin{table}[t]
\caption{Security profile of the valid candidate advisories ($n=469$) and of
the month-standardized reference ($n=13{,}199$). CWE shares are multi-label.
The last row is computed over the 468 candidates and the 12{,}951 reference
advisories that are on the two routes and have a source time no later than the
review (Section~\ref{sec:rq3})}
\label{tab:security}
\centering
\begin{tabular}{lrrr}
\toprule
Measure & Count & Candidates & Reference \\
\midrule
Critical severity & 49 & 10.4\% & 10.9\% \\
High severity & 200 & 42.6\% & 39.4\% \\
Medium severity & 193 & 41.2\% & 41.5\% \\
Low severity & 27 & 5.8\% & 8.2\% \\
\midrule
CVSS v3 observed & 350 & 74.6\% & 72.4\% \\
CVSS v4 only & 112 & 23.9\% & 26.5\% \\
EPSS observed & 369 & 78.7\% & 82.1\% \\
\midrule
CWE-918 (server-side request forgery) & 52 & 11.1\% & 5.4\% \\
CWE-22 (path traversal) & 50 & 10.7\% & 7.4\% \\
CWE-863 (incorrect authorization) & 32 & 6.8\% & 6.0\% \\
CWE-862 (missing authorization) & 29 & 6.2\% & 5.4\% \\
CWE-94 (code injection) & 25 & 5.3\% & 3.5\% \\
CWE-306 (missing authentication) & 19 & 4.1\% & 1.8\% \\
CWE-79 (cross-site scripting) & 18 & 3.8\% & 7.6\% \\
\midrule
GRA-linked & 386 & 82.3\% & 79.6\% \\
Reviewed within 24 hours of first disclosure & 77 & 16.5\% & 20.9\% \\
\bottomrule
\end{tabular}
\end{table}

The weakness profile departs more from the reference.  The candidates span 139
distinct CWE identifiers.  The most frequent are CWE-918 (server-side request
forgery, 52 advisories; 11.1\% versus 5.4\% in the reference), CWE-22 (path
traversal, 50; 10.7\% versus 7.4\%), CWE-863 (incorrect authorization, 32;
6.8\% versus 6.0\%), CWE-862 (missing authorization, 29; 6.2\% versus 5.4\%),
and CWE-94 (code injection, 25; 5.3\% versus 3.5\%).  CWE-306 (missing
authentication for a critical function) is more than twice as frequent as in
the reference (4.1\% versus 1.8\%), whereas cross-site scripting (CWE-79),
one of the most frequent weaknesses in the reference, is half as frequent
(3.8\% versus 7.6\%).  These differences are not driven by a single project:
the SSRF advisories come from 28 repositories and the path-traversal advisories
from 40, and the largest repository contributes 5.8\% of all candidates (the
five largest, 18.8\%).  They may nonetheless reflect the kinds of projects that
adopted coding agents early rather than the agents themselves; we did not
classify project domains.  Given that an advisory can contain multiple CWE
identifiers, these counts are multi-label.

Since the linkage follows the commits that advisories cite, this profile
describes the vulnerabilities whose remediation carries public agent
attribution; it does not measure the security quality of agent-written code.
Advisories usually cite the fix itself, but their references can also include
backports, tests, or release commits.  We did not classify the role of each
cited commit, so an attribution may concern a change adjacent to the fix rather
than the fix itself.

\begin{resultbox}
Approximately half of the valid candidates are high or critical severity (53.1\%,
versus 50.3\% in the month-standardized reference), and their severity, score,
and exploitability profile is close to that of other advisories published in
the same months.  Their weakness profile differs: server-side request forgery,
path traversal, code injection, and missing authentication are
over-represented, and cross-site scripting is under-represented.
\end{resultbox}

\subsection{RQ3: Disclosure route and review timing}
\label{sec:rq3}

Of the 469 valid candidates, 386 are GRA-linked, 82 are NVD-first, and one follows neither route. We set the singleton aside, leaving 468 candidates on the two substantive paths, 82.5\% of them GRA-linked. We measure review timing using the \emph{review lag}: the time, in days, between the publication of an advisory through the source associated with its route into the GHSA and its review by GitHub. For GRA-linked advisories, this timestamp is the publication date of the repository advisory; for NVD-first advisories, it is the NVD publication date. These timestamps identify the publication associated with each advisory's route into GHSA and do not necessarily represent the vulnerability's earliest public disclosure. The same definition applies to every group (Section~\ref{sec:paths}).   We compare the candidates
with two reference groups drawn from the same snapshot, both excluding
candidates and advisories that lack a source time or whose source time follows
the review (12 contemporaneous and 72 historical advisories).  The
\emph{contemporaneous} reference contains the 12{,}951 advisories published in
the months that contain candidates, month-standardized as in
Section~\ref{sec:rq2}.  The \emph{historical} reference contains the 10{,}403
advisories published from July~2022, after the NVD backfill, to August~2025,
the period covered by our earlier study \citep{segal2026ghsa}.

Route composition depends on the reference.  Against the historical
reference, candidates are far more concentrated in the GRA path (82.5\% versus
44.3\%).  Against advisories published in the same months, the difference
nearly vanishes (82.5\% versus 80.1\%): by 2026, when most candidates were
published, GRA had become the dominant route for all reviewed advisories.  The
historical contrast therefore reflects a platform-wide shift over time rather
than a property of AI-associated remediation.

\begin{figure}[t]
\centering
\input{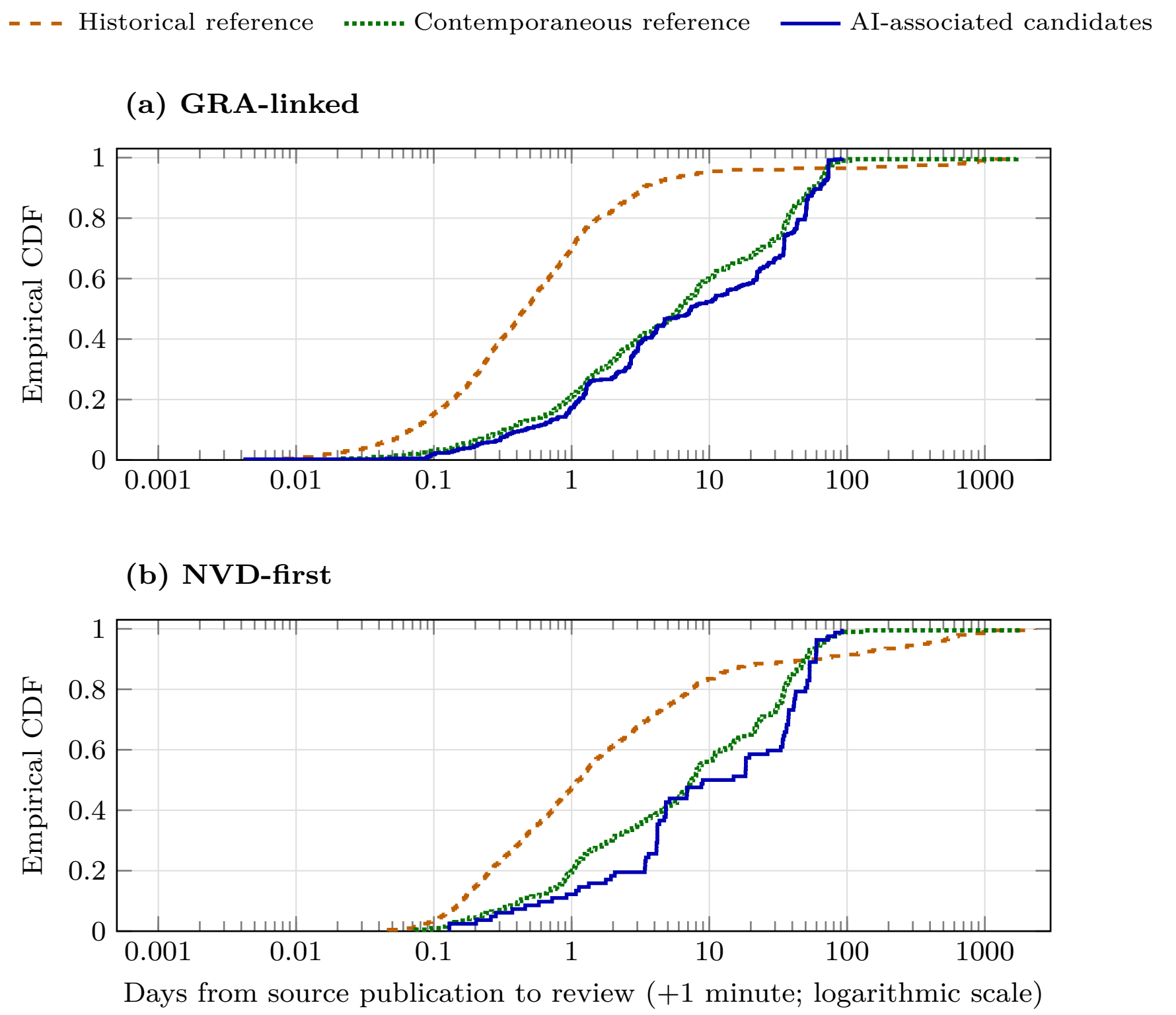}
\caption{Empirical CDFs of the review lag from source publication to GitHub
review on the GRA-linked route (a, $n=386$ candidates, $9{,}578$
contemporaneous, $4{,}612$ historical) and the NVD-first route (b, $n=82$
candidates, $3{,}373$ contemporaneous, $5{,}791$ historical).  The
contemporaneous reference is month-standardized.  One minute is added only to
display very short lags on the logarithmic axis; the reference curves are
plotted at 0.5-percentage-point resolution.}
\label{fig:review-cdfs}
\end{figure}

Within each route, the candidates resemble the contemporaneous reference and
differ sharply from the historical one (Fig.~\ref{fig:review-cdfs}).  On the
GRA-linked route the median lag is 7.4 days for the candidates, 6.0 days in the
contemporaneous reference, and 0.4 days in the historical one; the shares
reviewed within one day are 17.4\%, 21.2\%, and 69.9\%.  On the NVD-first route
the medians are 12.0, 7.4, and 1.1 days, and the shares within one day are
12.2\%, 19.8\%, and 47.5\%.  A candidate advisory is reviewed sooner than a
contemporaneous one in 45.8\% of GRA-linked pairs (95\% interval
40.6\%--51.0\%) and in 42.5\% of NVD-first pairs (34.6\%--51.1\%), so neither
difference is clear.  Against the historical reference the same probabilities
are 16.3\% (13.1\%--20.4\%) and 25.5\% (20.1\%--33.2\%).  The historical
reference thus describes an earlier, faster period.  Its NVD-first route
nevertheless has by far the heaviest extreme tail: the 95th percentile is
338.0 days, compared with 59.8 days for the candidates and 63.6 days for the
contemporaneous reference, while on the GRA-linked route the historical 95th
percentile (8.2 days) is far below the candidates' (73.1 days) and the
contemporaneous reference's (69.1 days).

\begin{figure}[t]
\centering
\input{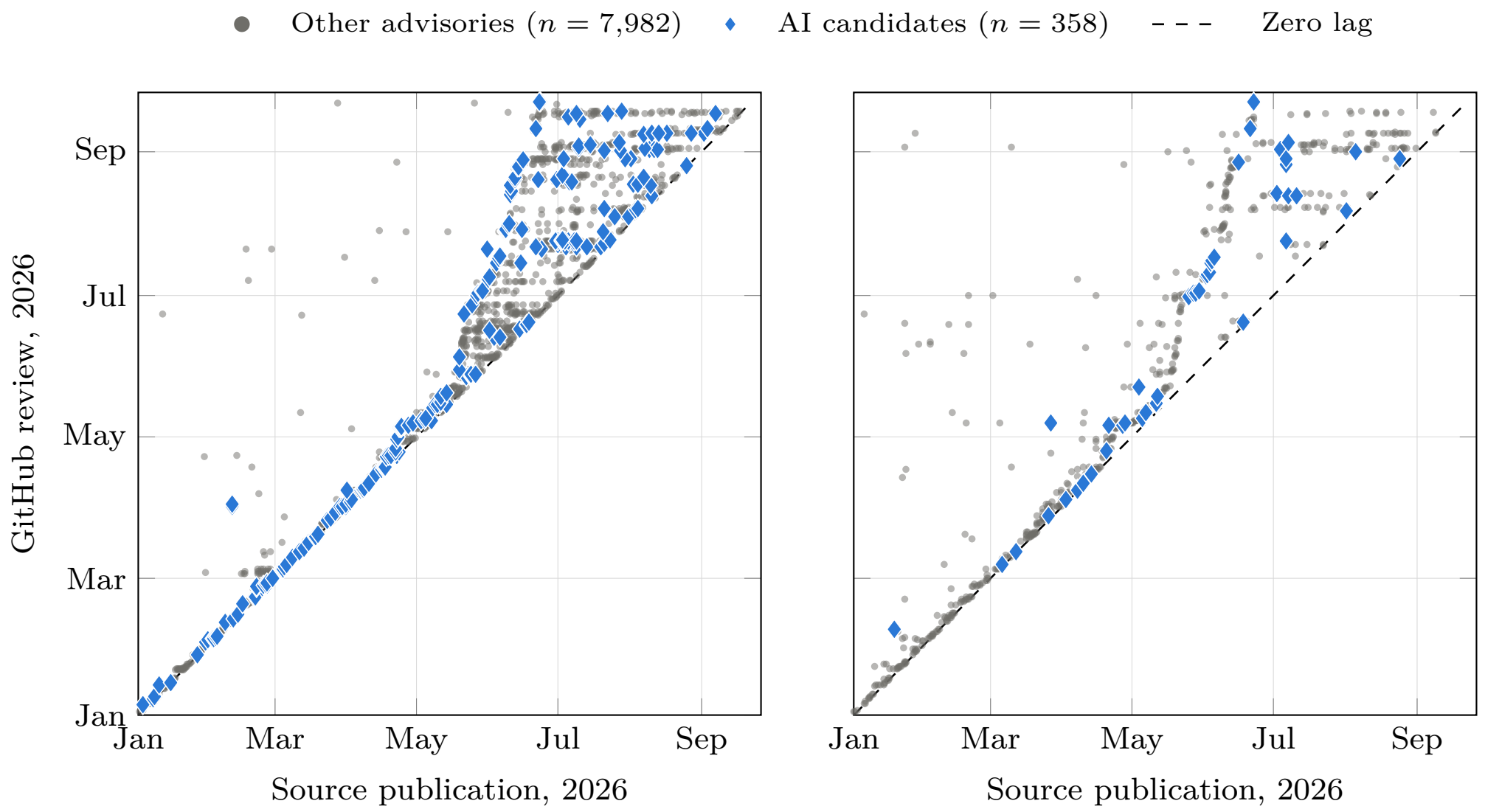}
\caption{Date of GitHub's review against the date of the source publication
for advisories published in 2026 on the GRA-linked route (a) and the NVD-first
route (b).  The vertical distance to the dashed diagonal is the review lag.
Panel (a) shows 358 AI-associated candidates and 7{,}982 other advisories;
panel (b) shows 73 candidates and 1{,}976 other advisories.  Candidates are
drawn over the other advisories. } 
\label{fig:review-order}
\end{figure}

The lags in Fig.~\ref{fig:review-cdfs} aggregate a period in which the review
queue changed.  Figure~\ref{fig:review-order} plots, for the advisories
published in 2026, the date of GitHub's review against the date of the source
publication.  Until about April, the points lie on the diagonal: advisories were
reviewed almost as soon as they were disclosed.  From May onward, on both
routes, review dates fall progressively behind publication dates.  Reviews
proceed approximately in publication order (first in, first out), so the points
bend away from the diagonal along a single curve, and they occur in bursts: a
horizontal streak is one review date that covers advisories published over
several weeks.  The candidates lie on the same curve as the other advisories and
inside the same bursts; neither route shows a separate fast lane or a separate
slow lane for them.  Section~\ref{sec:comparison-scope} returns to this
calendar trend.

Because the lag of an advisory depends mainly on when it joined this queue, and
it changed quickly within a month (for non-candidate GRA-linked advisories the
median rose from 0.4 days in January to 5.7 days in May and 29.3 days in July;
Section~\ref{sec:comparison-scope}), a month-standardized reference leaves part
of the calendar confound in place.  We therefore repeat the comparison with a
\emph{week-standardized} reference.  It is built like the contemporaneous
reference, but within each route and with each source week in place of each
month: the non-candidates of a route are reweighted so that every source week
has the share it has among that route's candidates.  All 386 GRA-linked and all
82 NVD-first candidates have at least ten reference advisories in their week.
On the GRA-linked route the candidates' median lag is 7.4 days against 5.3 days
for this reference, 17.4\% of the candidates are reviewed within one day against
23.3\%, and a candidate is reviewed sooner than a reference advisory in 43.5\%
of pairs (95\% interval 38.4\%--47.9\%).  The gap is modest, but unlike the
month-standardized comparison its interval excludes one half: matched on the
week, GRA-linked candidates are slower than their peers, not faster.  On the
NVD-first route the medians are 12.0 and 8.5 days, the shares within one day are
12.2\% and 10.5\%, and the probability is 50.7\% (44.2\%--55.4\%), so the
candidates are indistinguishable from their peers.  These comparisons use only
advisories that had been reviewed at collection. 


The overall mean review lag reflects both the proportions of advisories
following each route and the mean lag within each route. Advisories using
coding agents average 21.95 days across the two routes, compared with 19.45
days in the contemporaneous reference and 35.23 days in the historical
reference. These means are also sensitive to long lags, particularly in the
historical group. Section~\ref{sec:model} decomposes the differences from both
reference groups into route-composition and within-route components.

\begin{resultbox}
Most candidates enter through the GRA path (82.5\%), but so do other
advisories published in the same months (80.1\%).  The large gap relative to
the historical cohort (44.3\%) reflects the platform's shift toward repository
advisories over time.  Review
lags lengthened across the whole platform in 2026 as reviews fell behind
publications, and the candidates sit in the same queue as other advisories.  On
the NVD-first route candidates are not clearly faster or slower than
contemporaneous advisories; on the GRA-linked route they are not faster, and
they are modestly slower than advisories published in the same week.  They look
slower than historical ones because review lags were far shorter in the
historical period.
\end{resultbox}
 
\section{A Route--Timing Decomposition}
\label{sec:model}


Review latency is jointly determined by \emph{where an advisory enters the
disclosure pipeline} and \emph{how long it takes to progress within that
pipeline}.  These mechanisms are substantively different.  Agent-associated
remediation may be more likely to use GitHub's repository-advisory workflow
without changing review speed within either route.  Conversely, automation may
change a route-specific handoff while leaving route choice unchanged.  

Accounting for  the two considered paths, the
AI-associated candidates have a mean review lag of 21.95 days, compared with
19.45 days in the contemporaneous reference and 35.23 days in the historical
one.  The groups differ in how often they take the GRA-linked path (82.5\%,
80.1\%, and 44.3\%) and in how long each route takes: GRA-linked means are
21.65, 19.67, and 23.79 days, and NVD-first means are 23.35, 18.59, and 44.34
days. In what follows, we further analyze the  mix of route choices and   within-route timings.


\subsection{Cohorts, paths, and outcome}
\label{sec:paths}

The AI-associated candidates correspond to  469 validated advisories identified by our
mining pipeline.  For the route analysis we retain the 468 on the GRA-linked
(386) and NVD-first (82) paths and set the singleton residual advisory aside,
as in Section~\ref{sec:rq3}.  Both comparison groups are drawn from the
\texttt{advisories} table of the same snapshot, which holds every reviewed
GHSA. 
We call an advisory a non-candidate when it is not among the 469 valid
candidates; the 29 rejected linkages therefore belong to the references.
Absence from the candidate list is not proof that no AI tool was involved.
Like the whole analysis, the comparison is restricted to reviewed  advisories. 

The \emph{contemporaneous} reference $C$ contains the 12{,}951 non-candidates
(9{,}578 GRA-linked, 3{,}373 NVD-first) published in the 19 months that
contain at least one candidate.  Each month is reweighted to the share it has
among the 468 candidates, split evenly across that month's advisories, so
that $C$ has the candidates' calendar mix.  The \emph{historical} reference $H$
contains the 10{,}403 non-candidates (4{,}612 GRA-linked, 5{,}791 NVD-first)
published from July~2022, after the NVD backfill transition, to
August~2025, the period covered by the earlier GHSA study
\citep{segal2026ghsa}; it is unweighted.  Both references exclude advisories
without a source time or with a source time later than the review (4 and 8 in
$C$, 6 and 66 in $H$), and we write $A$ for the AI-associated candidates.

\begin{figure*}[t]
\centering
\resizebox{0.97\textwidth}{!}{\begin{tikzpicture}[
  x=1cm,y=1cm,
  box/.style={draw=black!70, rounded corners=1.5pt, fill=white,
    minimum height=9mm, text width=22mm, align=center, font=\small},
  src/.style={box, line width=1pt},
  route/.style={font=\small\bfseries, align=right, text width=20mm},
  arr/.style={-{Latex[length=2mm]}, line width=.65pt, draw=black!75},
  dim/.style={{Latex[length=1.6mm]}-{Latex[length=1.6mm]}, line width=.6pt, draw=black!75},
  measure/.style={font=\scriptsize, fill=white, inner sep=1.5pt},
  note/.style={font=\scriptsize, align=center, text width=80mm}
]
\node[route, text=blue!65!black] at (0,1.3)
  {GRA-linked\\$A:386$\\$C:9{,}578$\\$H:4{,}612$};
\node[box] (gfix) at (2.4,1.3) {Referenced security-fix commit};
\node[src] (gra) at (5.85,1.3) {Repository advisory (GRA) published};
\node[box] (grev) at (9.3,1.3) {Published and reviewed GHSA};
\draw[arr] (gfix) -- (gra);
\draw[arr] (gra) -- (grev);
\draw[dim] (5.85,.55) -- node[measure] {$L$} (9.3,.55);
\node[note, text=blue!65!black] at (7.575,.1)
  {$\widetilde L_A=7.4$, $\widetilde L_C=6.0$, $\widetilde L_H=0.4$ days};

\node[route, text=orange!75!black] at (0,-1.55)
  {NVD-first\\$A:82$\\$C:3{,}373$\\$H:5{,}791$};
\node[box] (nfix) at (2.4,-1.55) {Referenced security-fix commit};
\node[src] (nvd) at (5.05,-1.55) {NVD record published};
\node[box] (npub) at (7.7,-1.55) {GHSA publication};
\node[box] (nrev) at (10.35,-1.55) {GitHub review};
\draw[arr] (nfix) -- (nvd);
\draw[arr] (nvd) -- (npub);
\draw[arr] (npub) -- (nrev);
\draw[dim] (5.05,-2.3) -- node[measure] {$L$} (10.35,-2.3);
\node[note, text=orange!75!black] at (7.7,-2.75)
  {$\widetilde L_A=12.0$, $\widetilde L_C=7.4$, $\widetilde L_H=1.1$ days};

\draw[black!20, line width=.4pt] (-1.1,-.2) -- (11.65,-.2);
\end{tikzpicture}}
\caption{The two disclosure paths retained in the comparative analysis.
$A$ denotes AI-associated candidates, $C$ the contemporaneous reference, and
$H$ the historical reference; $\widetilde L$ denotes the route-specific
(weighted) median review lag in days, measured from the source publication
(repository advisory or NVD record) to GitHub's review. } 
\label{fig:paths}
\end{figure*}
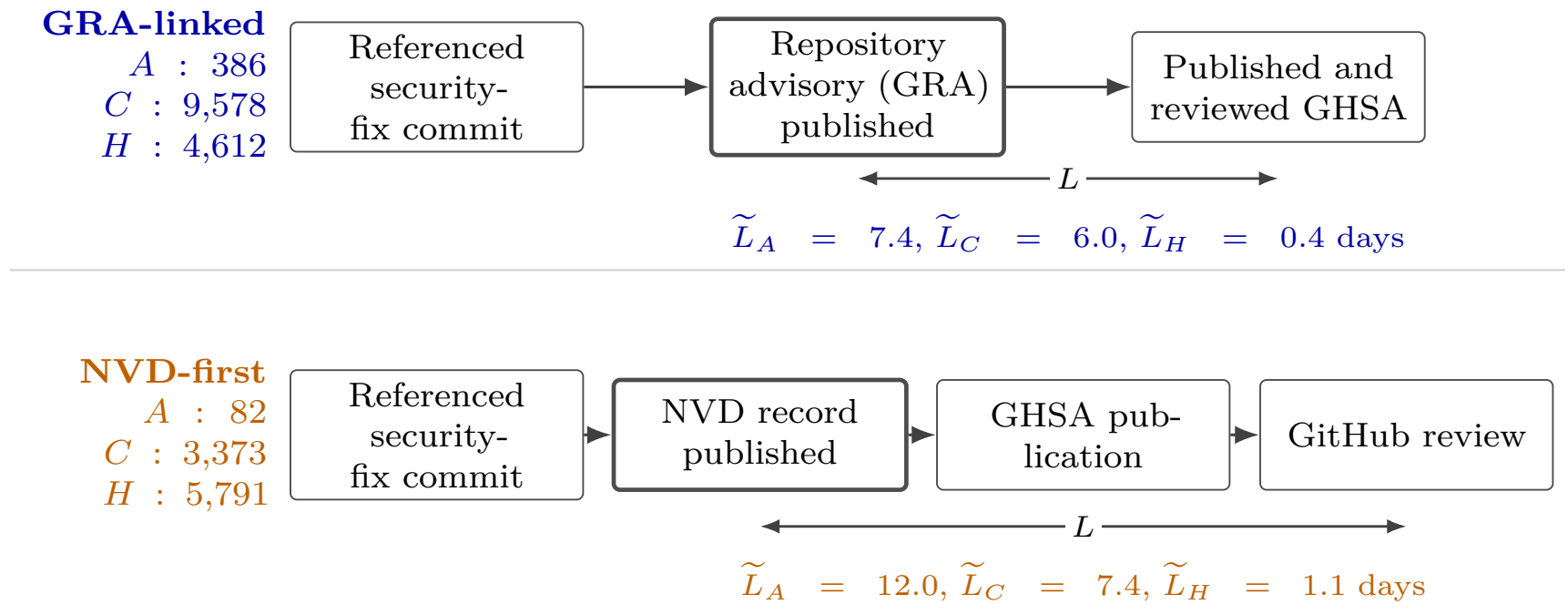

A GRA-linked advisory contains a repository-advisory URL.  An NVD-first
advisory lacks that link and contains an NVD publication timestamp.  The
two-route restriction places the singleton candidate residual category aside
and avoids assigning a common mechanism to heterogeneous remaining sources.
Figure~\ref{fig:paths} summarizes the observable paths and group sizes.

We measure the outcome from the source publication:
\begin{equation}
L=t^{\mathrm{review}}-t^{\mathrm{source}},
\label{eq:harmonized-lag}
\end{equation}
where $t^{\mathrm{review}}$ is GitHub's review time.  
Here, $t^{\mathrm{source}}$ denotes the start of the review-lag clock: the
publication time of the GitHub Repository Advisory for a GRA-linked advisory,
or the publication date recorded by the NVD for an NVD-first advisory. Thus,
the clock begins when the advisory becomes public through the source associated
with its route into the GHSA review pipeline, corresponding to the source-to-review
clock of~\cite{segal2026ghsa}.

\begin{table}[t]
\caption{A plausible pathway from coding-agent use to review demand.
The mechanisms are interpretive and were not directly measured in this
study.}
\label{tab:ai-review-demand}
\centering
\begin{tabularx}{\linewidth}{@{}Y Y Y@{}}
\toprule
Potential change & Effect on review demand & Possible consequence \\
\midrule
Lower cost of preparing a change &
More proposed changes, including redundant or unsolicited ones &
More work enters the review queue \\
Faster patch production &
Submissions arrive in a shorter period &
Temporary congestion if review capacity does not grow at the same rate \\
Limited project context or unclear rationale &
More effort to assess relevance, design fit, and correctness &
Longer triage or review for some changes \\
Triage policies and contribution controls &
Some proposals are filtered, redirected, or delayed before formal review &
Review demand may be contained, depending on policy effectiveness \\
\bottomrule
\end{tabularx}
\end{table}

\subsection{The impact of AI-associated candidates}

The AI-associated candidates have longer median review lags than the same-month
reference within both disclosure routes: 7.4 versus 6.0 days on the
GRA-linked route, and 12.0 versus 7.4 days on the NVD-first route.
Their mean lag is 2.49 days higher, although the confidence interval for this
difference spans zero, and their route shares are nearly identical to
those of the same-month reference. The candidates were reviewed sooner
than their same-month counterparts in 45.8\% of GRA-linked comparisons
and 42.5\% of NVD-first comparisons. Thus, the medians are higher for
the candidates in both routes, while their review-time distributions
substantially overlap with those of the same-month reference.

Review lags also increased during 2026 for GRA-linked advisories
generally: the median interval from repository-advisory publication to
the appearance of the reviewed GHSA rose from 0.4 days in January to
5.7 days in May and 41.4 days in September. The candidates exhibit the
same broad temporal pattern.

Table~\ref{tab:route-lags} shows the median review lag for contemporaneous and historical advisories. 
One possible mechanism 
that explains the recent increase in median review lag compared to the historical reference  
is that coding agents lower the cost of preparing
and submitting changes, allowing the number of proposed changes to grow
faster than maintainers' capacity to assess them. Faster production at the
submission stage need not mean faster review: maintainers still need to
establish whether a change is relevant, understand its context, check its
correctness and tests, and decide whether it fits the project's design.
If submissions arrive faster than they can be triaged and reviewed, the
queue of reviews may grow and typical waiting times may increase.


\subsection{Route-specific numerical comparison}
\label{sec:candidate-anatomy}

Table~\ref{tab:route-lags} reports the route-specific distributions of the
three groups.  The GRA-linked share of the candidates (82.5\%) is close to
that of $C$ (80.1\%) and 38.2 percentage points above that of $H$ (44.3\%).
On both routes the candidates' lags are long, and the contemporaneous
reference is the closest comparison.

\begin{table}[t]
\caption{Review lag (source publication to review) by group and disclosure
route.  The residual route is excluded.  Counts are unweighted; shares, means,
and quantiles use the group weights (the contemporaneous reference is
month-weighted).  Timing columns are in days.}
\label{tab:route-lags}
\centering
 \setlength{\tabcolsep}{3pt}
\begin{tabular}{lrrrrrrrr}
\toprule
Route & $n$ & Share & Mean & P25 & Median & P75 & P90 & $\leq 1$ day \\
\midrule
\multicolumn{9}{l}{\textit{AI-associated candidates ($A$)}} \\
GRA-linked & 386 & 82.5\% & 21.65 & 1.32 & 7.35 & 39.33 & 64.12 & 17.4\% \\
NVD-first  &  82 & 17.5\% & 23.35 & 3.62 & 11.98 & 41.24 & 59.38 & 12.2\% \\
\addlinespace
\multicolumn{9}{l}{\textit{Contemporaneous reference ($C$)}} \\
GRA-linked & 9{,}578 & 80.1\% & 19.67 & 1.25 & 6.00 & 34.00 & 56.01 & 21.2\% \\
NVD-first  & 3{,}373 & 19.9\% & 18.59 & 1.23 & 7.42 & 31.23 & 50.07 & 19.8\% \\
\addlinespace
\multicolumn{9}{l}{\textit{Historical reference ($H$)}} \\
GRA-linked & 4{,}612 & 44.3\% & 23.79 & 0.18 & 0.45 & 1.18 & 3.22 & 69.9\% \\
NVD-first  & 5{,}791 & 55.7\% & 44.34 & 0.33 & 1.10 & 5.10 & 55.28 & 47.5\% \\
\bottomrule
\end{tabular}
\end{table}

The group means are reconstructed, up to rounding, by their route mixtures:
\begin{align}
\mu_A &=0.825(21.65)+0.175(23.35)=21.95\ \text{days}, \\
\mu_C &=0.801(19.67)+0.199(18.59)=19.45\ \text{days}, \\
\mu_H &=0.443(23.79)+0.557(44.34)=35.23\ \text{days}.
\label{eq:observed-mixtures}
\end{align}
The historical reference differs from the others in both location and tail.
Its GRA-linked and NVD-first medians (0.45 and 1.10 days) are a small fraction
of the candidates' (7.35 and 11.98 days), and 69.9\% and 47.5\% of its
advisories are reviewed within one day, versus 17.4\% and 12.2\% of the
candidates.  Yet its NVD-first mean (44.34 days) is almost twice the
candidates' (23.35 days), because of an extreme tail: the longest lag on either
route is 91.9 days among the candidates, 1{,}852.1 days in $C$, and 2{,}319.1
days in $H$.  As reported in Section~\ref{sec:rq3}, a candidate
advisory is reviewed sooner than a contemporaneous one in 45.8\% of GRA-linked
and 42.5\% of NVD-first pairs, and sooner than a historical one in 16.3\% and
25.5\%.  Figure~\ref{fig:review-cdfs} shows both the central shift and the
tails.

\subsection{Exact mean decomposition}
\label{sec:decomposition}

Next, we decompose the difference in mean review lag between advisories using AI 
coding agents and each reference group into two components: the contribution
associated with differences in the proportions following the GRA-linked and
NVD-first routes, and the contribution associated with differences in mean
review lag within each route. 

Let  $Z$ denote the set of advisories of interest, where $Z\in\{A,B\}$. Here,  $A$ refers to the  AI-associate  candidates and $B$ refers to the baseline reference group, $B\in\{C,H\}$, where $C$ and $H$ refer to the contemporaneous and historical advisories, respectively. Let $R\in\mathcal{R}=\{G,N\}$ denote the GRA-linked and NVD-first routes.
With the group weights of Section~\ref{sec:paths}, define
\begin{equation}
q_{z,r}=\Pr(R=r\mid Z=z),\qquad
\mu_{z,r}=\mathbb{E}[L\mid Z=z,R=r].
\label{eq:components}
\end{equation}
The group mean is the finite mixture
\begin{equation}
\mu_z=\mathbb{E}[L\mid Z=z]
=\sum_{r\in\mathcal{R}}q_{z,r}\mu_{z,r}.
\label{eq:mixture}
\end{equation}


The mean review lag difference $\Delta=\mu_A-\mu_B$ can be decomposed as
\begin{equation}
\Delta=
\underbrace{\sum_{r\in\mathcal{R}}
(q_{A,r}-q_{B,r})\mu_{B,r}}_{\Delta_{\mathrm{route}}}
+
\underbrace{\sum_{r\in\mathcal{R}}
q_{A,r}(\mu_{A,r}-\mu_{B,r})}_{\Delta_{\mathrm{within}}}.
\label{eq:decomposition}
\end{equation}
The first term changes route composition while retaining the reference's
route-specific timings; the second changes route-specific timings while
retaining the candidate route mix, and splits into one contribution per route.

Indeed, to verify the above equality note that 
\begin{align*}
\mu_A-\mu_B
&=\sum_r q_{A,r}\mu_{A,r}-\sum_r q_{B,r}\mu_{B,r}\\
&=\left(\sum_r q_{A,r}\mu_{B,r}-\sum_r q_{B,r}\mu_{B,r}\right)
 +\left(\sum_r q_{A,r}\mu_{A,r}-\sum_r q_{A,r}\mu_{B,r}\right),
\end{align*}
which yields Eq.~\ref{eq:decomposition}.

\begin{figure}[t]
\centering
\resizebox{\textwidth}{!}{\begin{tikzpicture}[
  stage/.style={draw=black!70, rounded corners=2pt, fill=white,
    text width=52mm, minimum height=12mm, align=center, font=\small},
  component/.style={draw=#1!70!black, rounded corners=2pt, fill=#1!7,
    text width=52mm, minimum height=21mm, align=center, font=\small},
  head/.style={font=\small\bfseries, text width=52mm, align=center},
  arr/.style={-{Latex[length=2mm]}, line width=.75pt, draw=black!70}
]
\node[head] (hc) at (0,0) {Contemporaneous\\reference ($B=C$)};
\node[stage, below=4mm of hc] (c) {\textbf{Reference mean}\\
  $\mu_C=19.45$ days \quad [16.38,\ 22.60]};
\node[component=blue, below=5mm of c] (croute) {\textbf{Route composition}\\
  change $q_C\rightarrow q_A$; keep reference timings\\
  $\Delta_{\mathrm{route}}=+0.03$ days\\
  \relax[$-0.19,\ 0.31$]};
\node[stage, below=5mm of croute] (cstd) {\textbf{Route-standardized intermediate}\\
  $\sum_r q_{A,r}\mu_{C,r}=19.48$ days};
\node[component=orange, below=5mm of cstd] (cwithin) {\textbf{Within-route timing}\\
  keep $q_A$; change $\mu_C\rightarrow\mu_A$\\
  $\Delta_{\mathrm{within}}=+2.47$ days\\
  \relax[$-1.75,\ 7.15$]};
\node[stage, below=5mm of cwithin] (ca) {\textbf{AI-associated mean}\\
  $\mu_A=21.95$ days \quad [16.78,\ 27.19]};
\draw[arr] (c) -- (croute);
\draw[arr] (croute) -- (cstd);
\draw[arr] (cstd) -- (cwithin);
\draw[arr] (cwithin) -- (ca);
\node[head] (hh) at (6.2,0) {Historical\\reference ($B=H$)};
\node[stage, below=4mm of hh] (h) {\textbf{Reference mean}\\
  $\mu_H=35.23$ days \quad [26.53,\ 46.76]};
\node[component=blue, below=5mm of h] (hroute) {\textbf{Route composition}\\
  change $q_H\rightarrow q_A$; keep reference timings\\
  $\Delta_{\mathrm{route}}=-7.84$ days\\
  \relax[$-16.57,\ -1.33$]};
\node[stage, below=5mm of hroute] (hstd) {\textbf{Route-standardized intermediate}\\
  $\sum_r q_{A,r}\mu_{H,r}=27.39$ days};
\node[component=orange, below=5mm of hstd] (hwithin) {\textbf{Within-route timing}\\
  keep $q_A$; change $\mu_H\rightarrow\mu_A$\\
  $\Delta_{\mathrm{within}}=-5.45$ days\\
  \relax[$-14.24,\ 2.92$]};
\node[stage, below=5mm of hwithin] (ha) {\textbf{AI-associated mean}\\
  $\mu_A=21.95$ days \quad [16.23,\ 27.33]};
\draw[arr] (h) -- (hroute);
\draw[arr] (hroute) -- (hstd);
\draw[arr] (hstd) -- (hwithin);
\draw[arr] (hwithin) -- (ha);
\end{tikzpicture}}
\caption{Numerical route--timing decomposition against the contemporaneous
(left) and historical (right) references.  The intermediate mean applies the
AI-associated route shares to the reference's route-specific means.  Brackets
show 95\% repository-cluster bootstrap intervals.}
\label{fig:decomposition}
\end{figure}
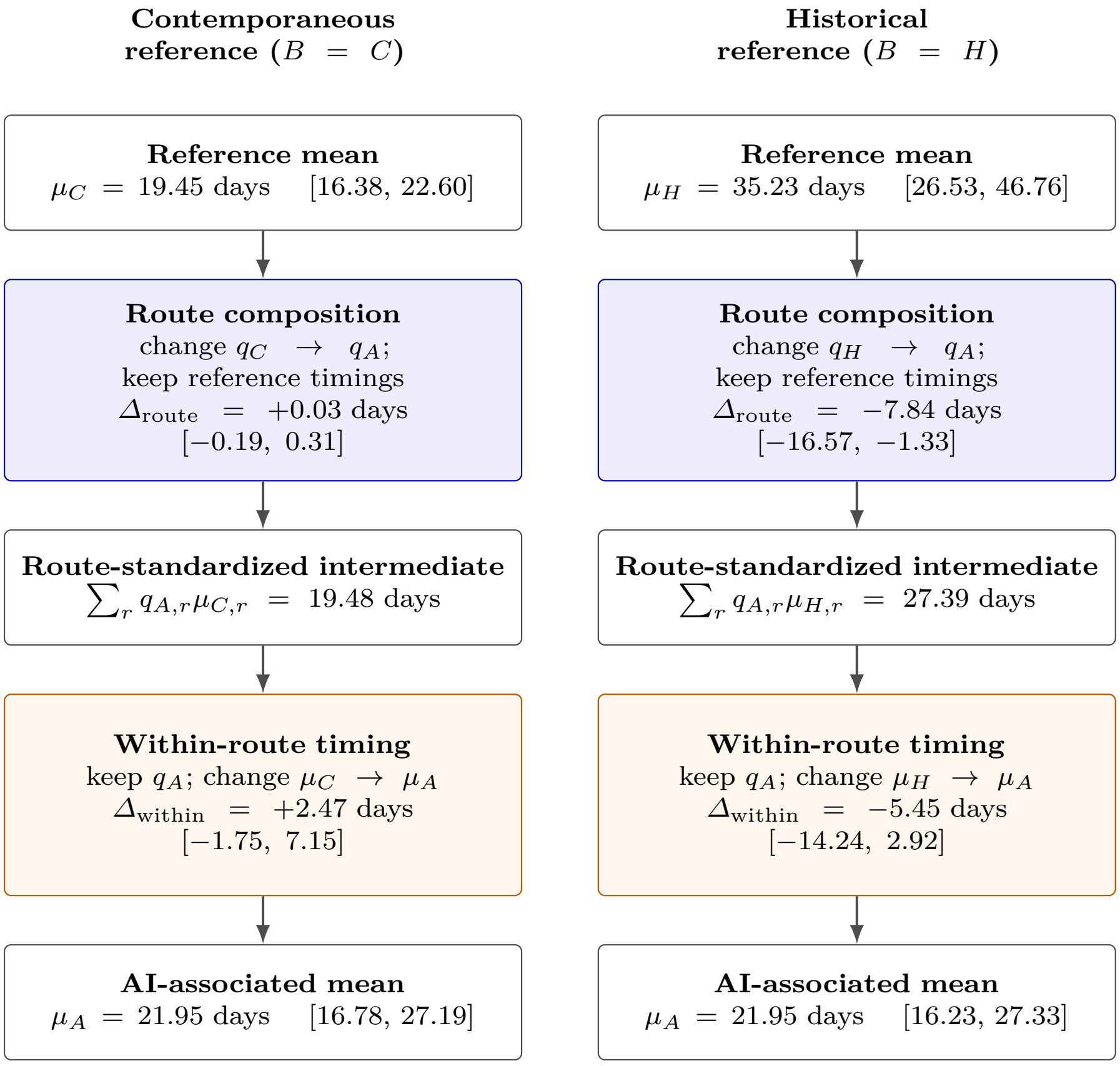

Table~\ref{tab:decomposition-results} and Fig.~\ref{fig:decomposition} report
the estimates against both references.  Confidence intervals come from 1{,}000
bootstrap replicates that resample repositories independently within each group,
preserving within-project dependence; they use the resampling of
Section~\ref{sec:rq3}, including month weights recomputed from the resampled
candidates for $C$.  Each comparison has its own resamples, so the interval of
$\mu_A$ differs slightly between the two columns.

\begin{table}[t]
\caption{Route--timing decomposition of the review lag in days against each
reference group.  Brackets show 95\% repository-cluster bootstrap intervals.}
\label{tab:decomposition-results}
\centering
\scriptsize
\setlength{\tabcolsep}{3pt}
\begin{tabular}{lrlrl}
\toprule
 & \multicolumn{2}{c}{Contemporaneous ($B=C$)} & \multicolumn{2}{c}{Historical ($B=H$)} \\
\cmidrule(lr){2-3}\cmidrule(lr){4-5}
Quantity & Estimate & 95\% interval & Estimate & 95\% interval \\
\midrule
Reference mean, $\mu_B$ & 19.45 & [16.38,\ 22.60] & 35.23 & [26.53,\ 46.76] \\
AI-associated mean, $\mu_A$ & 21.95 & [16.78,\ 27.19] & 21.95 & [16.23,\ 27.33] \\
Marginal difference, $\Delta$ & $+2.49$ & [$-1.72$,\ 7.20] & $-13.28$ & [$-25.36$,\ $-2.90$] \\
Route composition, $\Delta_{\mathrm{route}}$ & $+0.03$ & [$-0.19$,\ 0.31] & $-7.84$ & [$-16.57$,\ $-1.33$] \\
Within-route timing, $\Delta_{\mathrm{within}}$ & $+2.47$ & [$-1.75$,\ 7.15] & $-5.44$ & [$-14.24$,\ 2.92] \\
\quad GRA-linked contribution & $+1.63$ & [$-2.63$,\ 6.08] & $-1.77$ & [$-10.17$,\ 5.98] \\
\quad NVD-first contribution & $+0.83$ & [$-0.33$,\ 2.13] & $-3.68$ & [$-7.81$,\ $-0.91$] \\
\bottomrule
\end{tabular}
\end{table}

Against the contemporaneous reference the candidates' mean is 2.49 days higher,
with an interval that spans zero.  Their route mix is essentially that of $C$
($\Delta_{\mathrm{route}}=+0.03$), so the whole difference is within-route
timing, mostly on the GRA-linked route, and it is not distinguishable from
zero.  Against the historical reference the candidates are 13.28 days
faster on the marginal mean, and 7.84 days (59.0\%) of that difference is route
composition: replacing the historical shares (44.3\% GRA, 55.7\% NVD-first)
with the candidate shares (82.5\% GRA, 17.5\% NVD-first) while holding the
historical route-specific means fixed lowers the expected lag from 35.23 to
27.39 days.  The within-route component is $-5.44$ days; its confidence  interval includes
zero, and its sign depends on the historical tail, as further detailed next. 

\subsection{Distributional interpretation}
\label{sec:distributional-standardization}

Next, we extend our analysis beyond mean and median review lags, to account for the  distribution of the review lag. 
  Let
$S_{z,r}(t)=\Pr(L>t\mid Z=z,R=r)$ be the route-specific survival function.
Then
\begin{equation}
S_z(t)=\sum_{r\in\mathcal{R}}q_{z,r}S_{z,r}(t).
\label{eq:survivalmix}
\end{equation}
A route-standardized candidate curve under the reference's route mix is
\begin{equation}
S_A^{\mathrm{std}}(t)=
\sum_{r\in\mathcal{R}}q_{B,r}S_{A,r}(t).
\label{eq:standardized}
\end{equation}
Comparing $S_A^{\mathrm{std}}$ with $S_B$ removes the marginal route-share
difference while retaining within-route distributional differences.  Under the
historical route mix, the standardized candidate curve has 14.5\% of
advisories reviewed within one day, against 57.5\% for $H$, and 37.2\% with a
lag above 30 days, against 7.7\%.  Under the contemporaneous mix the gaps are
smaller but point the same way: 16.3\% versus 20.9\% within one day, and 34.8\%
versus 26.8\% above 30 days.  Thus, within routes, the bulk of the AI   candidates'
distribution is slower than both contemporaneous and historical references.

The mean points the other way against $H$ because it is dominated by lags that the AI-associated  
candidates cannot match: the longest  AI-associated candidate review lag is
91.9 days, while $H$ reaches 2{,}319.1.  As a sensitivity check, we cap every
lag at 91.9 days, the longest AI-associated candidate lag; the cap is 
used   to expose the role of the tail.  Against $H$, the capped historical
mean falls from 35.23 to 8.25 days, the route component remains negative
($-2.85$ days, interval $[-4.40,\ -1.69]$), and the within-route component
changes sign to $+16.55$ days ($[10.84,\ 22.25]$), driven mainly by the
GRA-linked route ($+14.48$ days, $[8.86,\ 20.42]$); the capped marginal
difference is $+13.69$ days ($[7.97,\ 19.27]$).  Against $C$ the capped
within-route component is $+4.03$ days ($[-0.08,\ 8.29]$) and the capped marginal
difference is $+4.03$ days ($[-0.08,\ 8.23]$).  The negative within-route mean
against $H$ is therefore produced by the extreme historical tail rather than
by faster typical review of the AI-associated   candidates.  


\subsection{Interpretation, calendar trend, and scope}
\label{sec:comparison-scope}

Relative to the
historical post-backfill data, AI-associated candidates have a lower marginal
mean review lag primarily because they are much more likely to travel
through the GRA-linked path.   
Relative to advisories published in the same
months, they have no clearly different mean, and their route mix is the same.
Neither comparison supports the claim that AI accelerates review within a fixed
route: within routes the median review lag is larger than in both references.

The historical contrast is confounded with calendar time, because review lags
lengthened substantially across the whole platform.  Among all valid
non-candidate GRA-linked advisories, the median lag is 0.4 days for advisories
published in January 2026, 5.7 days in May, 29.3 days in July, and 41.4 days in
September; the candidates' medians for the same months are 0.5, 4.5, 34.6, and
32.4 days.  Because 431 of the 469 valid candidates were published in 2026,
month standardization is essential, which is why the contemporaneous reference
is the primary comparison. 


As a second sensitivity analysis we repeat the decomposition on the same groups
with the interval from GHSA publication to review.  GRA-linked lags are then
zero, and only the NVD-first terms remain.  The marginal difference is
$+0.47$ days against $C$ ($[-1.69,\ 2.94]$), with route composition $-0.43$
($[-1.84,\ 1.23]$) and within-route timing $+0.90$ ($[-0.27,\ 2.16]$); against
$H$ it is $-16.94$ days ($[-29.06,\ -8.28]$), of which 85.0\% is route
composition ($-14.40$ days, $[-23.49,\ -8.06]$) and the within-route component
is $-2.53$ days ($[-6.55,\ 0.24]$).  The qualitative conclusion is unchanged,
but the GHSA clock understates the lag of the GRA-linked route and, with it,
the weight of the calendar trend.

\section{Discussion}
\label{sec:discussion}

\subsection{Evidence and scale of AI coding agents in GHSAs} \label{sec:evid}

Our   results evidence the  existence and scale of AI coding agents in GHSAs: hundreds of reviewed advisories
now reference commits whose public Git metadata explicitly attributes work to
coding agents.  The evidence spans general-purpose agents and security-specific
automation such as Copilot Autofix.  


The evidence does not establish how much code an agent produced, whether a
human substantially rewrote it, or whether the final change was correct because
of the agent. Indeed, co-author trailers are simple declarations.   Exact bot
identities are stronger operational evidence but still do not reveal the
division of labor.  Instruction files and configuration directories are weaker
still: they show that maintainers prepared a repository for agent interaction,
not that a particular security fix was agent-authored.

The timing result is equally specific.  AI-associated candidates have a lower marginal mean lag than the
historical cohort (13.28 days), and 59\% of that difference is their much
greater use of the GRA path.  The rest, a negative within-route component, comes
from the historical extreme tail and reverses sign when lags are capped.
Against advisories published in the same months, the candidates have the same
route mix and no clearly different mean ($+2.49$ days, interval spanning zero).
Within each route, advisories using coding agents have a higher median review
lag than advisories in either reference group. On the GRA-linked route, the
median is 7.4 days, compared with 6.0 days in the same-month reference and
0.4 days in the historical reference; on the NVD-first route, it is 12.0 days,
compared with 7.4 and 1.1 days, respectively. These data therefore do not
support claims that AI-assisted review is faster.

\textbf{Throughput constraint. }
The more striking pattern is platform-wide.  The median lag between the
publication of a repository advisory and its reviewed GHSA rose from 0.4 days
for advisories published in January 2026 to 5.7 days in May and 41.4 days in
September, and the AI-associated candidates track it.  Public timestamps alone cannot identify the cause, but GitHub has
reported that inbound advisory volume rose sharply in 2026 (1{,}560 reviewed
advisories in May alone, over five times its typical monthly output) and that
review times for new advisories lengthened as a result
\citep{GitHubBlogAdvisoryVolume2026}.  The direction of our measured trend is
consistent with this account of a throughput constraint.

\subsection{Implications for maintainers and platforms}

Security advisories would benefit from structured provenance.  Today,
attribution is scattered across free-form trailers, bot identities, pull
request text, and repository files.  A structured field distinguishing
generation, review, testing, and documentation roles would reduce ambiguity
and enable more reliable audits.  Such provenance should identify the human
maintainer who accepted the change and should not imply that an automated tool
bears responsibility for disclosure decisions.

Maintainers can improve traceability by preserving agent attribution through
squash merges, requiring security tests for agent-assisted fixes, and recording
whether the agent generated the patch, suggested it, reviewed it, or merely
edited supporting text.  For high-impact vulnerabilities, repository policy
could require a human-authored rationale explaining the vulnerable behavior,
the security invariant restored by the patch, and the regression tests that
exercise~it.

For GitHub and other advisory platforms, the concentration of candidate
observations in GRAs, which mirrors that of all advisories published in the
same months, suggests that repository-native disclosure is the natural place to
capture provenance.  Optional structured metadata could support
research and governance without exposing private prompts or confidential draft
advisories.

\section{Threats to Validity}
\label{sec:threats}

\paragraph{Construct validity.}
Git author, committer, and co-author fields are unauthenticated.  An exact
trailer can be spoofed, omitted, or added by a human after the agent's work.
Conversely, an agent contribution may leave no detectable signal.  Our rules
favor precision, so the corpus is a lower-bound view of visible attribution,
not an estimate of all agent participation.  The identified Devin human-name
collision demonstrates that even seemingly distinctive tool names require
validation.  Artifact-only candidates are reported separately because they do
not support commit-level authorship.

\paragraph{Link validity.}
An advisory reference is not guaranteed to identify the unique patch commit.
References may point to a release, test, merge, backport, documentation change,
or related discussion.  One commit can support multiple advisories, and one
advisory can cite multiple commits.  We therefore report all counting units and
avoid equating an advisory--commit link with vulnerability introduction.
Diff-level classification is needed for a stronger remediation-role taxonomy.

\paragraph{Internal validity.}
The analysis is observational.  Repository maturity, activity, ecosystem,
security policy, project popularity, and calendar time may jointly influence
agent use, GRA adoption, and review latency.  In particular, 431 of 469 valid
candidates were published in 2026, whereas the historical comparison ends in
August~2025, and review lags lengthened markedly in between.  The
decomposition reveals whether a marginal difference is associated with route
shares or within-route means, and the same-month reference removes the
calendar-month confound, but neither removes project-composition confounding or
identifies causal effects.

\paragraph{Timing validity.}
The review lag starts at the source publication, dated by the
repository-advisory publication or the NVD publication; private coordination
before that moment is not observable.  The interval from GHSA publication to
review gives the same qualitative conclusions but is exactly zero for
GRA-linked advisories, which hides the delay we measure.  The analysis covers
only advisories that GitHub had reviewed at collection, so long lags are less
observable for recent publications; the same-month reference shares this
truncation by construction, but the historical one does not. 

\paragraph{External validity.}
The study includes public GitHub repositories linked to reviewed GHSAs.
Private repositories, unreviewed advisories, proprietary incident systems, and
agents that do not preserve public attribution are outside scope.  Tool-family
frequencies reflect both use and observability; they should not be generalized
to overall market share or relative security quality.

\paragraph{Reliability.}
Scans are pinned to immutable targets and
findings use stable identities.  Nevertheless, the ecosystem evolves rapidly:
new agent accounts, trailer conventions, and configuration paths will require
rule updates.  We will publish the rule catalog, snapshot date, analysis code,
and candidate table so that later replications can distinguish true temporal
change from detector evolution.

\section{Conclusion}
\label{sec:conclusion}

We presented a security-advisory-centered study of coding agents in real
software repositories.  By joining high-precision local-Git findings to commits
referenced by reviewed GHSAs, we obtained 469 valid candidate advisories across
189 repositories after removing 29 advisories affected by human-name
collisions.  Direct attribution appears in 448 advisories, while 21 are
supported   by repository instruction or configuration artifacts.  The
candidates cover a broad range of severity and weakness classes, and most are
connected to the fast review, repository-native GRA path, as are other advisories
published in the same months.

The main methodological lesson is that attribution must be layered.  A bot
identity, an explicit co-author trailer, and an \tool{AGENTS.md} file are not
interchangeable evidence.  The main analytical lesson is that GHSA route composition between the NVD and GRA paths
must be separated from within-route timing before differences in review latency
are interpreted. In addition, we observed interesting synchronization effects: dating a GRA-linked advisory from
its GHSA publication yields a negligible review lag, as publication and review typically occur concomitantly for GRA-first advisories,  whereas dating it from the
repository advisory exposes a delay that grew across the platform during 2026.
Measured from source publication, the AI-associated candidates' mean review lag is 13.28 days
lower than the historical data and 2.49 days higher than that of advisories
published in the same months.  Of the historical difference, 7.84 days come from
their 82.5\% GRA share versus 44.3\% historically, and the within-route
component depends on the historical extreme tail; against same-month advisories
the route mix is the same and the difference is not clear.  Within each route,
the AI-associated candidates are not faster, suggesting that the increase in the reported inbound advisory volume and throughput constraints may play a role in  the review delays. 

Future work should connect repository-level evidence to the full remediation
and review process. Linking agent-attributed commits to pull requests could
allow researchers to measure submission volume, time to first review, review
duration, and time to merge, while distinguishing direct agent contributions
from repository-level adoption signals. Longitudinal data could then test
whether changes in agent use are associated with changes in review workload
and delay, accounting for route, publication month, and advisory
characteristics. Public traces now make coding-agent involvement in security
remediation measurable; our work is a first step towards a research agenda that aims to understand  what happens throughout
the review process and  to establish  whether involvement of  AI coding agents   leads to faster
or more effective remediation.

 \section{Replication and Data Availability}
\label{sec:replication}

The replication package   includes a  versioned   catalog of detectors of AI coding agents, schema,
notebooks, deduplicated candidate table, validation decisions, and scripts that
produce the tables and figures. It is archived on Zenodo at \url{https://doi.org/10.5281/zenodo.23378198}.
 \appendix

\section{Detection-rule catalog}
\label{app:catalog}

Table~\ref{tab:catalog} lists the values recognized by each detection rule
summarized in Table~\ref{tab:rules}, together with the implementation
identifier.

\begin{table}[!ht]
\caption{Recognized values per detection rule.}
\label{tab:catalog}
\centering
\footnotesize
\begingroup\sloppy
\begin{tabularx}{\textwidth}{@{}p{3.3cm}Y@{}}
\toprule
Rule   & Recognized values \\
\midrule
Message signal \newline
(\path{commit_ai_evidence.message-signal}) &
Line-anchored \tool{Co-authored-by: Name <email>} trailers whose name is an
allowlisted bot handle, contains one of the whole tokens \tool{copilot},
\tool{claude}, \tool{codex}, \tool{chatgpt}, \tool{gemini}, \tool{cursor},
\tool{devin}, \tool{aider}, \tool{coderabbit}, \tool{sweep}, \tool{codeium},
\tool{sourcery}, or \tool{codeflash}, or whose address is a canonical Copilot,
Anthropic, or OpenAI address.  Explicit generation phrases (e.g., ``generated
by GitHub Copilot,'' ``Claude Code created,'' ``AI-generated'').  Workflow
forms (a standalone ``Apply Copilot suggestion'' line; ``Copilot
Autofix''). \\
\addlinespace
Known bot identity \newline
(\path{commit_bot_author.known-ai-account}) &
Author or committer name equal, case-insensitively, to
\tool{github-copilot[bot]}, \tool{copilot-swe-agent[bot]},
\tool{devin-ai-integration[bot]}, \tool{sweep-ai[bot]},
\tool{coderabbitai[bot]}, \tool{sourcery-ai[bot]}, or
\tool{codeflash-ai[bot]}.  E-mail addresses are not used to infer identity. \\
\addlinespace
Instruction file \newline
(\path{instruction_file.universal}) &
Addition of, or rename into, one of the root files \tool{AGENTS.md},
\tool{CLAUDE.md}, \tool{CONTEXT.md}, \tool{GEMINI.md}, \tool{CODEX.md},
\tool{COPILOT.md}, \tool{CURSOR.md}, \tool{CLINE.md}, \tool{QWEN.md}, or
\tool{JULES.md} (exact, case-sensitive).  For a rename, the previous path is
retained. \\
\addlinespace
Agent configuration \newline
(\path{agent_config_file.universal}) &
Addition of, or rename into, one of the root files \tool{.cursorrules},
\tool{.clinerules}, \tool{.aider.conf.yml}, \tool{.aiderignore},
\tool{.aider.model.settings.yml}, \tool{opencode.json}, or
\tool{opencode.jsonc}; or addition of any file below one of the top-level
directories \tool{.claude}, \tool{.cursor}, \tool{.gemini}, \tool{.codex},
\tool{.windsurf}, \tool{.amazonq}, \tool{.continue}, \tool{.opencode}, or
\tool{.qwen}, recorded at the directory level. \\
\bottomrule
\end{tabularx}
\endgroup
\end{table}

\section{Examples of project contribution policies discussed in response
to AI-generated submissions}
\label{app:project-policies}

Public discussion of AI-assisted contributions includes projects that
have restricted outside pull requests or introduced additional
conditions for accepting them (see Table~\ref{tab:project-contribution-policies}). These examples provide context for the
possibility that lower-cost change production can increase the demand
placed on human review (see Section~\ref{sec:evid}). 

Reported policies differ. tldraw and Flue have been described as
automatically closing external pull requests and redirecting proposals
to issues or discussions. Ladybird announced that it would stop
accepting public pull requests. Ghostty, by contrast, has been reported
to permit AI-assisted pull requests under conditions such as prior
agreement on an issue, while rejecting unsolicited submissions. Some restrict all external
pull requests, whereas others restrict only certain AI-assisted or
unsolicited contributions. Project policies also evolve, so the examples are a snapshot of public reporting.

\begin{table}[t]
\caption{Illustrative project policies concerning outside or
AI-assisted contributions. Policies may change over time.}
\label{tab:project-contribution-policies}
\centering
\begin{tabularx}{\linewidth}{@{}p{0.17\linewidth}Y Y@{}}
\toprule
Project & Reported policy & Interpretation \\
\midrule
Ladybird &
The contribution guide states that pull requests from non-maintainers
are not accepted. &
Restriction on outside pull requests; the policy itself does not establish
that AI was its sole cause. \\

tldraw &
Public reporting describes automatic closure of external pull requests
and redirection of proposals to issues or discussions. &
Restriction on the external PR channel. \\

Flue &
Public reporting describes automatic closure of external pull requests,
with proposals redirected to issues or discussions. &
Restriction on the external PR channel. \\

Ghostty &
The contribution guide asks that pull requests be associated with a
previously accepted issue, and its AI policy sets conditions for
AI-assisted contributions. &
Conditional contribution policy, rather than a blanket ban on pull
requests. \\
\bottomrule
\end{tabularx}
\end{table}

\section*{Declarations}

\paragraph{Funding}
This work was partially supported by INES.IA (National Institute of Science and Technology for Software Engineering Based on and for Artificial Intelligence), www.ines.org.br, CNPq grant 408817/2024-0. This work was also partially supported by CAPES and FAPERJ under grants E-26/204.268/2024 and E-26/260.168/2026, CNPq grants 444956/2024-7, 424622/2021-1 and 315106/2023-9, as well as Finep PlatCiber.

\paragraph{Ethical Approval}
The study analyzes public software repositories and public security-advisory
metadata and does not involve human-subject interaction.  Evidence is reported
in aggregate except for public repository identifiers already contained in the
replication dataset.

\paragraph{Informed Consent}
 Not applicable.

\paragraph{Conflict of Interest}
The authors declare that they have no known competing financial interests or personal relationships that could have appeared to influence the work reported in this paper.

\paragraph{Clinical Trial Number}
Not applicable.

\paragraph{Author Contributions.}
Claudio Segal and Paulo Segal contributed data curation, software, and
investigation.  Hudson Silva Borges   contributed  methodology, data collection,
validation   and interpretation.  Daniel S.\ Menasch\'e contributed
conceptualization, analytical modeling, supervision, and writing. Paulo Silveira, Eduardo de Almeida, Felipe Paixão, Joanna C.S. Santos    and Raphael Machado  contributed writing and mentoring. Anton Kocheturov and  Gaurav Srivastava contributed   industrial insights. All authors
reviewed and approved the manuscript.

\bibliographystyle{spbasic}
\bibliography{references}

@misc{GitHubAdvisoryDatabase,
  title        = {GitHub Advisory Database: Security vulnerability database inclusive of CVEs and GitHub-originated security advisories},
  author       = {{GitHub}},
  year         = {2025},
  url          = {https://github.com/advisories},
  note         = {See also \url{https://github.com/github/advisory-database} and \url{https://docs.github.com/en/code-security/security-advisories/working-with-repository-security-advisories/about-repository-security-advisories}. Accessed: 2025-08-18}
}

@inproceedings{li2017large,
  title={A large-scale empirical study of security patches},
  author={Li, Frank and Paxson, Vern},
  booktitle={Proceedings of the 2017 ACM SIGSAC Conference on Computer and Communications Security},
  pages={2201--2215},
  year={2017}
}

@article{ponta2020detection,
  title={Detection, assessment and mitigation of vulnerabilities in open source dependencies},
  author={Ponta, Serena Elisa and Plate, Henrik and Sabetta, Antonino},
  journal={Empirical Software Engineering},
  volume={25},
  number={5},
  pages={3175--3215},
  year={2020},
  publisher={Springer}
}

@article{liu2025empirical,
  title={An empirical study on vulnerability disclosure management of open source software systems},
  author={Liu, Shuhan and Zhou, Jiayuan and Hu, Xing and Cogo, Filipe Roseiro and Xia, Xin and Yang, Xiaohu},
  journal={ACM Transactions on Software Engineering and Methodology},
  year={2025},
  publisher={ACM New York, NY}
}

@inproceedings{decan2018impact,
  title={On the impact of security vulnerabilities in the npm package dependency network},
  author={Decan, Alexandre and Mens, Tom and Constantinou, Eleni},
  booktitle={Proceedings of the 15th international conference on mining software repositories},
  pages={181--191},
  year={2018}
}

@inproceedings{ayala2025mixed,
  title={{A Mixed-Methods  Study of Open-Source  Software Maintainers On Vulnerability Management and Platform Security Features}},
  author={Ayala, Jessy and Tung, Yu-Jye and Garcia, Joshua},
  booktitle={34th USENIX Security Symposium (USENIX Security 25)},
  pages={2105--2124},
  year={2025}
}

@inproceedings{ayala2025deep,
  title={A deep dive into how open-source project maintainers review and resolve bug bounty reports},
  author={Ayala, Jessy and Ngo, Steven and Garcia, Joshua},
  booktitle={2025 IEEE Symposium on Security and Privacy (SP)},
  pages={522--538},
  year={2025},
  organization={IEEE}
}

@inproceedings{alexopoulos2021vulnerability,
  title={{Who are vulnerability reporters? A large-scale empirical study on FLOSS}},
  author={Alexopoulos, Nikolaos and Meneely, Andrew and Arnouts, Dorian and M{\"u}hlh{\"a}user, Max},
  booktitle={Proceedings of the 15th ACM/IEEE international symposium on empirical software engineering and measurement (ESEM)},
  pages={1--12},
  year={2021}
}

@article{ayala2025investigating,
  title={Investigating Vulnerability Disclosures in Open-Source Software Using Bug Bounty Reports and Security Advisories},
  author={Ayala, Jessy and Tung, Yu-Jye and Garcia, Joshua},
  journal={arXiv preprint arXiv:2501.17748},
  year={2025}
}

@article{wang2017characterizing,
  title={Characterizing and modeling patching practices of industrial control systems},
  author={Wang, Brandon and Li, Xiaoye and de Aguiar, Leandro P and Menasche, Daniel S and Shafiq, Zubair},
  journal={Proceedings of the ACM on Measurement and Analysis of Computing Systems},
  volume={1},
  number={1},
  pages={1--23},
  year={2017},
  publisher={ACM New York, NY, USA}
}

@misc{AboutGithubAdvisoryDatabase,
    title  = {About the GitHub Advisory Database},
    author = {{GitHub Docs}},
    year   = {2025},
    url    = {https://docs.github.com/en/code-security/security-advisories/working-with-global-security-advisories-from-the-github-advisory-database/about-the-github-advisory-database},
    note   = {Accessed: 2025-10-16}
}

@inproceedings{bozorgi2010beyond,
  title={Beyond heuristics: learning to classify vulnerabilities and predict exploits},
  author={Bozorgi, Mehran and Saul, Lawrence K and Savage, Stefan and Voelker, Geoffrey M},
  booktitle={Proceedings of the 16th ACM SIGKDD international conference on Knowledge discovery and data mining},
  pages={105--114},
  year={2010}
}

@article{arora2010empirical,
  title={An empirical analysis of software vendors' patch release behavior: impact of vulnerability disclosure},
  author={Arora, Ashish and Krishnan, Ramayya and Telang, Rahul and Yang, Yubao},
  journal={Information Systems Research},
  volume={21},
  number={1},
  pages={115--132},
  year={2010},
  publisher={INFORMS}
}

@inproceedings{zimmermann2019small,
  title={Small world with high risks: A study of security threats in the npm ecosystem},
  author={Zimmermann, Markus and Staicu, Cristian-Alexandru and Tenny, Cam and Pradel, Michael},
  booktitle={28th USENIX Security symposium (USENIX security 19)},
  pages={995--1010},
  year={2019}
}

@article{allodi2017security,
  title={Security events and vulnerability data for cybersecurity risk estimation},
  author={Allodi, Luca and Massacci, Fabio},
  journal={Risk Analysis},
  volume={37},
  number={8},
  pages={1606--1627},
  year={2017},
  publisher={Wiley Online Library}
}

@article{he2023automating,
  title={{Automating dependency updates in practice: An exploratory study on GitHub Dependabot}},
  author={He, Runzhi and He, Hao and Zhang, Yuxia and Zhou, Minghui},
  journal={IEEE Transactions on Software Engineering},
  volume={49},
  number={8},
  pages={4004--4022},
  year={2023},
  publisher={IEEE}
}

@inproceedings{mirhosseini2017can,
  title={Can automated pull requests encourage software developers to upgrade out-of-date dependencies?},
  author={Mirhosseini, Samim and Parnin, Chris},
  booktitle={2017 32nd IEEE/ACM international conference on automated software engineering (ASE)},
  pages={84--94},
  year={2017},
  organization={IEEE}
}

@inproceedings{kinsman2021software,
  title={How do software developers use github actions to automate their workflows?},
  author={Kinsman, Timothy and Wessel, Mairieli and Gerosa, Marco A and Treude, Christoph},
  booktitle={2021 IEEE/ACM 18th International Conference on Mining Software Repositories (MSR)},
  pages={420--431},
  year={2021},
  organization={IEEE}
}

@misc{AboutDependabotAlets,
    title  = {About Dependabot alerts},
    author = {{GitHub Docs}},
    year   = {2025},
    url    = {https://docs.github.com/en/code-security/dependabot/dependabot-alerts/about-dependabot-alerts},
    note   = {Accessed: 2025-10-16}
}

@misc{GitHubBlogNVDBackfill,
  title        = {All historical NVD advisories are now listed on GitHub},
  author       = {{GitHub Blog}},
  year         = {2022},
  url          = {https://github.blog/changelog/2022-06-08-all-historical-nvd-advisories-are-now-listed-on-github/},
  note         = {Accessed: 2025-09-12}
}

@inproceedings{kancharoendee2025categorizing,
  title={On Categorizing Open Source Software Security Vulnerability Reporting Mechanisms on GitHub},
  author={Kancharoendee, Sushawapak and Phichitphanphong, Thanat and Jongyingyos, Chanikarn and Reid, Brittany and Kula, Raula Gaikovina and Choetkiertikul, Morakot and Ragkhitwetsagul, Chaiyong and Sunetnanta, Thanwadee},
  booktitle={2025 IEEE International Conference on Software Analysis, Evolution and Reengineering (SANER)},
  pages={751--756},
  year={2025},
  organization={IEEE}
}

@inproceedings{ayala2024poster,
  title={{Poster: A glimpse of vulnerability disclosure behaviors and practices using GitHub projects}},
  author={Ayala, Jessy and Tung, Yu-Jye and Garcia, Joshua},
  booktitle={Proceedings of the 45th IEEE Symposium on Security and Privacy (S\&P)},
  year={2024}
}

@article{li2025aidev,
  author  = {Li, Hao and Zhang, Haoxiang and Hassan, Ahmed E.},
  title   = {The Rise of {AI} Teammates in Software Engineering ({SE}) 3.0: Characterizing Agentic Coding in {GitHub} Pull Requests},
  journal = {arXiv preprint arXiv:2507.15003},
  year    = {2025},
  doi     = {10.48550/arXiv.2507.15003}
}

@inproceedings{segal2026ghsa,
  author    = {Segal, Claudio and Segal, Paulo and Banjar, Carlos Eduardo and Paix\~ao, Felipe de Sant'Anna and Borges, Hudson Silva and Silveira, Paulo and Almeida, Eduardo Santana de and Santos, Joanna C. S. and Kocheturov, Anton and Srivastava, Gaurav and Menasch\'e, Daniel S.},
  title     = {Characterizing and Modeling the {GitHub} Security Advisories Review Pipeline},
  booktitle = {Proceedings of the 23rd International Conference on Mining Software Repositories},
  pages     = {311--322},
  year      = {2026},
  doi       = {10.1145/3793302.3793360}
}

@misc{siddiq2026securityageaiteammates,
      title={{Security in the Age of AI Teammates: An Empirical Study of Agentic Pull Requests on GitHub}}, 
      author={Mohammed Latif Siddiq and Xinye Zhao and Vinicius Carvalho Lopes and Beatrice Casey and Joanna C. S. Santos},
      year={2026},
      eprint={2601.00477},
      archivePrefix={arXiv},
      primaryClass={cs.CR},
      url={https://arxiv.org/abs/2601.00477}, 
}

@article{pearce2025asleep,
  title={Asleep at the keyboard? assessing the security of github copilot’s code contributions},
  author={Pearce, Hammond and Ahmad, Baleegh and Tan, Benjamin and Dolan-Gavitt, Brendan and Karri, Ramesh},
  journal={Communications of the ACM},
  volume={68},
  number={2},
  pages={96--105},
  year={2025},
  publisher={ACM New York, NY, USA}
}

@misc{fu2025securityweaknessescopilotgeneratedcode,
      title={{Security Weaknesses of Copilot-Generated Code in GitHub Projects: An Empirical Study}}, 
      author={Yujia Fu and Peng Liang and Amjed Tahir and Zengyang Li and Mojtaba Shahin and Jiaxin Yu and Jinfu Chen},
      year={2025},
      eprint={2310.02059},
      archivePrefix={arXiv},
      primaryClass={cs.SE},
      url={https://arxiv.org/abs/2310.02059}, 
}

@inproceedings{rooijendijk2026said,
  author    = {Rooijendijk, Pien and Treude, Christoph and Wessel, Mairieli},
  title     = {Who Said {CVE}? How Vulnerability Identifiers Are Mentioned by Humans, Bots, and Agents in Pull Requests},
  booktitle = {Proceedings of the 23rd International Conference on Mining Software Repositories},
  year      = {2026},
  publisher = {ACM},
  address   = {New York, NY, USA},
  doi       = {10.1145/3793302.3793616},
  note      = {arXiv:2601.19636}
}

@misc{GitHubBlogAdvisoryVolume2026,
  title        = {Inside the {Advisory Database} and what happens when vulnerability volume breaks records},
  howpublished = {The GitHub Blog},
  year         = {2026},
  url          = {https://github.blog/security/supply-chain-security/},
  note         = {Accessed: 2026-10-15}
}

\end{document}